\documentclass{article}
\usepackage{arxiv}
\usepackage[utf8]{inputenc} 
\usepackage[T1]{fontenc}    
\usepackage{hyperref}       
\usepackage{url}            
\usepackage{booktabs}       
\usepackage{amsfonts}       
\usepackage{nicefrac}       
\usepackage{microtype}      
\usepackage{lipsum}		
\usepackage{graphicx}
\usepackage{doi}
\usepackage{longtable}
\usepackage{caption}
\usepackage{subcaption}
\usepackage{multirow}
\usepackage[font=small,labelfont=bf]{caption}

\title{A Test for Evaluating Performance in Human-Computer Systems}

\date{} 					

\author{ Andres Campero$^{a,*}$
	\And
	Michelle Vaccaro$^{a,*}$
	\And
	Jaeyoon Song$^a$
	\And
	Haoran Wen$^a$ 
	\And
	Abdullah Almaatouq$^a$ 
	\And
	Thomas Malone$^{a,\dagger}$ 
	\\ \\ 
	$^a$ Massachusetts Institute of Technology Center for Collective Intelligence, Cambridge, MA 02139 \\
	$^*$ Andres Campero and Michelle Vaccaro contributed equally to this work \\
	$^\dagger$ To whom correspondence may be addressed: email: malone@mit.edu
}

\begin{document}
\maketitle

\begin{abstract}
The Turing test for comparing computer performance to that of humans is well known, but, surprisingly, there is no widely used test for comparing how much better human-computer systems perform relative to humans alone, computers alone, or other baselines.  Here, we show how to perform such a test using the ratio of means as a measure of effect size. Then we demonstrate the use of this test in three ways. First, in an analysis of 79 recently published experimental results, we find that, surprisingly, over half of the studies find a \textit{decrease} in performance, the mean and median ratios of performance improvement are both approximately 1 (corresponding to no improvement at all), and the maximum ratio is 1.36 (a 36\% improvement). Second, we experimentally investigate whether a higher performance improvement ratio is obtained when 100 human programmers generate software using GPT-3, a massive, state-of-the-art AI system. In this case, we find a speed improvement ratio of 1.27 (a 27\% improvement). Finally, we find that 50 human \textit{non-}programmers using GPT-3 can perform the task about as well as--and less expensively than--the human programmers. In this case, neither the non-programmers nor the computer would have been able to perform the task alone, so this is an example of a very strong form of human-computer synergy.
\end{abstract}

\vspace{12pt}

\centerline
  {\large \bfseries \scshape Significance Statement}
  \begin{quote}
  
  The Turing test inspired generations of computer scientists to try to develop software as intelligent as humans. But we also need, not just more intelligent software, but more intelligent human-computer systems.  The test proposed here can help scientifically evaluate and, thus, spur progress in developing such systems. For instance, contests like those that helped develop autonomous vehicles might, in this case, stimulate competition among teams of computer and social scientists developing human-computer systems that perform far better than either people or computers alone. We believe this could benefit our economy and society by fostering the development of software to augment humans rather than just replace them. And in the long run, it might also lead to creating ``superintelligent'' human-computer systems.
  \end{quote}


\section{Introduction}
Starting in the 1950s, when Alan Turing proposed his famous ``Turing test,'' he inspired generations of computer scientists with the vision of developing artificially intelligent computers that would someday equal human performance~\cite{grosz2012question, machinery1950computing}. Substantial progress has been made in realizing this vision, and there have been many studies comparing computer performance to that of humans~\cite{tsividis2017human, silver2016mastering, lake2017building}. But no computer today can equal human performance in all ways, and many experts believe that such an accomplishment is still far in the future~\cite{brooks2017seven, lake2017building}.

In the meantime, people and computers working together in various ways have become commonplace in many aspects of modern life, and there have been a number of studies evaluating the performance of such human-computer combinations~\cite{tschandl2020human, vaccaro2019effects, groh2022deepfake}. But it is surprising that there is not yet a widely recognized scientific analog of the Turing test to systematically measure the improvements in how well people and computers together can perform tasks better than either could alone or better than some other relevant benchmark. Such a test could, for example, help  stimulate and guide a search for human-computer systems that perform far better than anything we have today.

Here we propose such a test and demonstrate its use in three studies. In the first study, we use it to systematically analyze and compare a number of recent studies of human-computer combinations. Then, in two original human subject studies, we use it to evaluate how people combined with a massive, state-of-the-art AI program can perform a typical software development task. We also discuss various potential uses of this test, including (a) measuring the specialized collective intelligence of human-computer systems, (b) stimulating progress with contests analogous to the autonomous vehicle contests sponsored by DARPA~\cite{buehler2009darpa, behringer2004darpa}, and (c) providing opportunities for collaboration between computer scientists and social scientists to develop and study not only new technologies themselves but also novel ways of combining humans and computers to use these technologies.

\section{Approach}
There are two simple ideas behind the test we propose:
\begin{enumerate}
\item Instead of viewing humans and computers as competitors in performing tasks, view them as collaborators.
\item Instead of viewing human performance as an upper bound, try to maximize the ratio of improvement of the human-computer system relative to some benchmark such as humans only, computers only, or current practice.
\end{enumerate}
To formalize these ideas, we let

\vspace{6pt}
\hspace{35pt} $X_i$ = the average performance of a system of type $i$ on a given task

\hspace{35pt} $\rho = \frac{X_i}{X_j}$ 

where the values of $i$ and $j$  represent different types of systems such as:

\hspace{35pt}\textit{H = human}

\hspace{35pt}\textit{C = computer}

\hspace{35pt}\textit{HC = human-computer}

\hspace{35pt}\textit{B = placeholder for any relevant baseline}

\vspace{6pt}
With this formulation, when $\rho>1$, the human-computer combination outperforms the baseline, and when $\rho>>1$ there is substantial benefit.  We are particularly interested in maximizing a version of $\rho$ that measures the synergy~\cite{larson2010search, almaatouq2021task} between humans and computers, which we define as:

\vspace{6pt}
\hspace{35pt} $\hat{\rho}=\frac{X_{HC}}{max(X_H,X_C)}$
\vspace{6pt}

In this case, when $\hat{\rho}>1$, the combination of humans and computers performs better than either alone. In other words, when $\hat{\rho}>1$, there is synergy between the humans and computers.

Of course, to systematically measure any of these kinds of performance, we need to specify the task(s) being performed (e.g., recognizing faces, developing software), the performance dimension(s) being measured (e.g., speed, cost, quality), and details about the types of systems performing the tasks (such as capabilities of the humans and computers and configurations of the human-computer systems). For example, in specifying dimensions of performance to be measured, there are often multiple dimensions that can be traded off against each other (e.g., speed can often be increased by reducing quality). In applying this test, therefore, it is often useful to specify multi-dimensional performance measures such as maximizing speed subject to quality constraints.

It is important to note that these $\rho$ ratios constitute a measure of effect size called the ratio of means~\cite{Friedrich2008,GrissomKim2012ch2, Pustejovsky2018}. Like other effect size measures, this ratio quantifies the actual magnitude of an effect as opposed to, for example, t-tests or analyses of variance that quantify the probability a given result could have occurred by chance. Effect size measurements are usually more helpful in evaluating the practical significance of a result, not just its statistical significance, and they are increasingly expected in reports of scientific research (e.g., \cite{APA2020}).  

Two important benefits of this particular effect size measurement are: 
\begin{enumerate}
    \item \textit{Intuitive.} The ratio of means corresponds to the intuitive understanding of improvement as a factor or as a percentage -- a ratio of 2 indicates a doubling in performance relative to the baseline, and a ratio of 1.5 indicates a 50\% improvement. This relationship makes it easier for people, especially those without statistical training, to understand compared to other measures of effect size (such as standardized mean difference or correlation coefficients).
    \item \textit{Comparable.} Since the ratio of means is a unitless measure, it is comparable across many different units and scales of measurement. And unlike other measures of effect size, such as standardized mean difference, it does not depend on the number or distribution of results in a given sample, so it is comparable across many different experimental designs and populations. 
\end{enumerate}

To use this form of effect size measurement, the $X$'s usually need to be measured on a ratio scale with a ``real 0'' (see~\cite{Stevens1946,VellemanWilkinson1993} and SI Appendix Sec. S1.1). 

It is also helpful to transform measures where smaller values are desirable (such as time and cost) using what we call the \textit{desirable lower bound} transformation, which usually takes the form: $f(X) = \frac{1}{X}$. This makes such measures comparable to ones where larger values are better (such as speed and accuracy). For instance, this transformation converts time into its reciprocal, speed.

It is sometimes also helpful to use what we call the \textit{desirable upper bound} transformation to make values unbounded in the positive direction. For instance, in studies measuring accuracy in human-computer systems, the baseline performance of humans or computers alone may already be greater than 90\%. In these cases, without this transformation, the maximum improvement possible would be $\rho = \frac{1}{.90} = 1.05$. For more on these transformations, see SI Appendix Sec. S1.2.

In the remainder of this paper, we demonstrate the use of these $\rho$ ratios and transformations in three ways. We use them first to systematically analyze 79 previous experiments that compare the performance of human-computer combinations to humans alone and computers alone.  Since all these studies were published in 2021, they provide one representation of the current state-of-the-art in combining people and computers. We also illustrate the use of this approach in analyzing two original human subject experiments on a software development task performed by human-computer pairs. In both of these studies, the primary computer component is GPT-3~\cite{gpt3}, a massive, state-of-the-art AI system that uses machine learning techniques to produce text that is sometimes strikingly human-like. GPT-3 can also produce other kinds of text, such as the code for computer software, and that is the focus of the studies described here.

\section{Results}
\subsection{Study 1: Analysis of recent studies that evaluate human-computer performance}

People today increasingly look to computers---and especially artificial intelligence---to augment human intelligence in business~\cite{sanders2003efficacy}, medicine~\cite{longoni2019resistance}, government~\cite{valle2019review}, and many other aspects of life.  But how do we quantify the synergy in this type of collaboration?  Do human-computer teams actually achieve better outcomes than either working alone? And what magnitudes of improvement are found in recent research?  

To systematically address these questions, we first used the techniques of a systematic literature review~\cite{Kitchenham2004ProceduresFP, xiao2019guidance} to identify the relevant papers published in 2021 in a variety of computing- and social science-related outlets. In particular, we selected papers with one or more original experiments that evaluated computers helping humans perform tasks and that included quantitative measures of the performance of (1) the humans alone, (2) the computers alone, and (3) the human-computer systems. 

Using this selection process, we identified a total of 25 papers that met our inclusion criteria, and these papers included a total of 79 unique experimental results. Notably, most of the experimental results evaluate the performance of human-computer teams according to accuracy (63\%), but we find a total of 11 different metrics used across all the experiments.  These different units make comparisons of human-computer synergy across studies non-trivial --- one cannot simply compare an increase of, say, 10\% in accuracy, with a decrease of, say, 2 steps needed to complete the task. For more details on the methodology used and papers included, see SI Appendix Sec. 2.1.

\begin{figure*}[tbh]
\centering
\includegraphics[width=\textwidth]{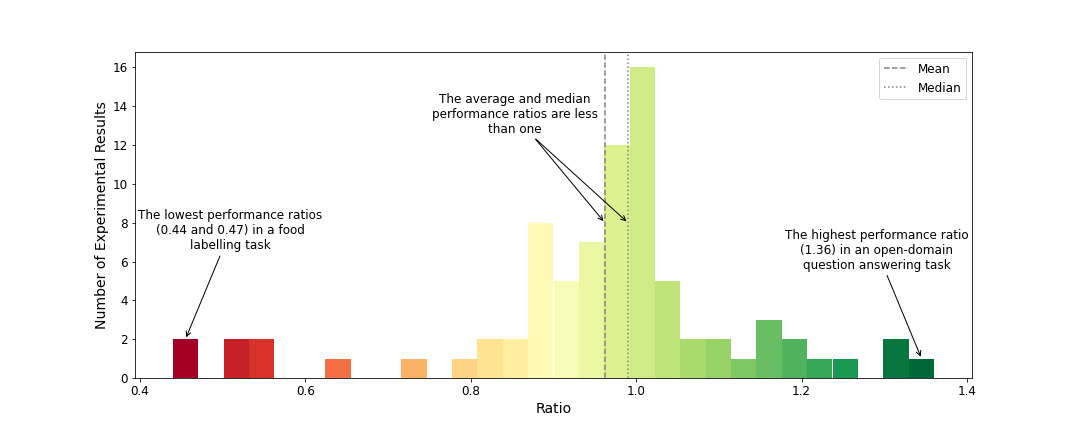}
\caption{The distribution of ratios from the experimental results in our review.  The values of the  ratios range from 0.44 to 1.36, and the mean and median are approximately 1 (mean = 0.96, median = 0.99). Only 38\% of these experimental measurements reflect evidence of positive synergy as indicated by a performance ratio greater than one.}
\label{fig:ratios_main}
\end{figure*}

To compare and summarize the different results in all these different studies, we calculated the $\hat{\rho}$ ratios as defined above. For metrics in which lower values correspond to better performance, we used the desirable lower bound transformation specified above before computing the ratios.   Figure~\ref{fig:ratios_main} displays the distribution of the results. The values of the  ratios range from 0.44 to 1.36, and the mean and median are approximately 1 (mean=0.96, median=0.99). In fact, only 38\% of these experimental measurements reflect evidence of positive synergy as indicated by  $\hat{\rho} > 1$. (See SI Appendix Table S1 for the values calculated in each case, and SI Appendix Sec. S2.1.5 for various robustness checks.)

In other words, despite the widespread talk of humans and computers dramatically amplifying each other's strengths~\cite{Wilson2018Collab}, we found that, in this sample at least, less than half of the combinations of humans and computers actually achieved performance improvements.

Most of the ratios less than 1.0 come from experiments in which the human-computer systems perform worse than the computer alone according to some measure of accuracy, sensitivity, or specificity~\cite{Skirzynski2021AutomaticDO, Kyriacou2021LearningTW, Buccinca2021ToTO}.  At the bottom of the distribution, for example, Buçinca et al conduct an experiment in which participants view an image of a meal and must substitute the highest carbohydrate ingredient with a low carbohydrate ingredient still similar in flavor~\cite{Buccinca2021ToTO}.  For this task, the AI algorithm achieves 75\% accuracy while the human-AI system only reaches 33\% accuracy, and, notably, the poor human-computer synergy occurs despite the cognitive forcing function the researchers designed to elicit an appropriate level of reliance on the AI tool~\cite{Buccinca2021ToTO}.

On the other hand, multiple studies do find evidence of synergistic human-computer systems.  For instance, in the study with the highest $\hat{\rho}$ ratio in our study, Gonzalez et al~\cite{Gonzalez2021DoEH} demonstrate how algorithms can improve human decision-making in open-domain question answering tasks.  In their experiment, the condition in which humans work alone achieves an accuracy of 57\% and the condition in which the algorithm works alone achieves an accuracy of 50\%.  But when humans view the model's prediction, confidence level, and a brief extractive explanation, they achieve an accuracy of 78\% and a $\hat{\rho}$ of 1.36.  The authors also find performance improvements when humans only see the model prediction or see different types of explanations, though of slightly lesser magnitudes (1.19 - 1.36).

Most studies, however, find that collaboration does not result in these large magnitudes of performance change --- the majority of the experiments in our review lead to ratios near one.

\subsubsection{Survey of expert opinion}
To contextualize our findings, we conducted a small online survey of researchers actively engaged in empirical studies of human-computer teams.  We asked the respondents to estimate three summary statistics about the performance of human-computer teams in studies published in 2021. Figure S4 includes the full text of the questions, and Figures S5 and S6 show the distribution of responses.  Given some extreme outliers in the responses, we highlight the median and interquartile ranges (IQRs) as opposed to the mean and standard deviations of the estimates for each question.

Most notably, participants overestimated the average ratio of improvement in the performance of human-computer teams relative to the performance of humans alone or computers alone.  Their median estimate was 1.2x (1.05 - 1.5 IQR), versus our finding of 0.96.  They also overestimated the greatest ratio of improvement with a median estimate of 2.0x (1.5 - 4.0x IQR), versus our finding of 1.4. 

The median response was approximately accurate, however, for the estimate of the percentage of experiments in which the human-computer team performed better (median of 50\% (20 - 80\% IQR) versus our finding of 38\%). Surprisingly, we observe a wide range of responses to all questions, suggesting that even experts in this realm do not currently share a common understanding of the current state of knowledge in the area. For more details about the methods and results of the survey, See SI Appendix Sec. S2.2.

\subsection{Study 2: Software generation by human programmers and GPT-3}

Study 1 suggests that it may be more difficult to obtain large values of $\hat{\rho}$ than many people expect. But none of the systems analyzed in Study 1 used the surprisingly powerful new generation of massive AI systems that are pre-trained on vast amounts of text~\cite{gpt3, Xuetal2022}.  Might this new generation of technology allow larger ratios of improvement? And regardless of the level of improvement, are there statistical ways to estimate a ratio of means while simultaneously controlling for many other factors?

In Study 2, we address both of these questions in the context of a software development task. The task for subjects in this study is to develop software code in HyperText Markup Language (HTML) that will replicate a web page. The web pages include images, links to external sites, buttons, and text of different sizes and forms (see SI Sec. S3.1 for screenshots).

The subjects in this study are human programmers with expertise in HTML coding. We consider two experimental conditions. In the control condition (human-only or ``H''), the human generates all the code using a simple text editor with no AI support. Of course, the text editor is a simple kind of computer support, but for simplicity, we consider this a ``human-only'' condition because the computer support here is trivial relative to that in the treatment condition.

\begin{figure}[hh]
\centering
\includegraphics[width=.45\textwidth]{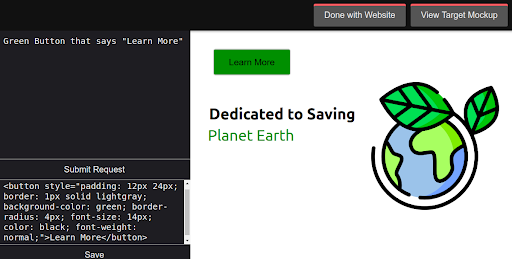}
\caption{Display of the interface for the Human-Computer condition. The top left corner is the field for the subjects to input the natural language instructions. Below on the left are the obtained HTML outputs which the user can modify or delete. On the right is the visual display of the rendered HTML items which can be dragged by the user to different positions. The figure illustrates some of the kinds of elements GPT-3 is able to create in response to simple textual descriptions, such as tables, buttons, images, and other HTML tags.}
\label{fig:interface}
\end{figure} 

In the treatment condition (human-computer or ``HC''), the human directs the high-level organization of the website using natural language while leaving the detailed code generation to the GPT-3 AI algorithm. The order and conditions in which a given subject sees the two tasks are randomized. For details about GPT-3 and its use here, see SI Sec. S3.2.

We do not test a computer-only (or ``C'') condition because we assume that neither GPT-3 nor any other currently available computer system is capable of performing alone the basic task in this study: looking at a purely visual representation of a web page and generating the HTML code needed to produce that web page. 

In other words, for Study 2, the conditions are:

\hspace{35pt}
\textit{H = human programmers with conventional HTML text editor}

\hspace{35pt}
\textit{C = computers (assumed impossible today, so not tested)}

\hspace{35pt}
\textit{HC = human programmers with GPT-3 interface}

As a performance measure, we focus on the speed of performing each task, subject to a constraint on the quality of the solutions. More specifically, we test the hypothesis that the combination of humans and computers (HC) can achieve synergy relative to both humans alone (H) and computers alone (C), subject to an acceptable quality constraint (>80\% correct submissions).  In other words, we hypothesize (pre-registered) that 

\hspace{35pt} $\hat{\rho}=\frac{X_{HC}}{max(X_H,X_C)}>1$

and

\hspace{35pt} $A_{HC},\thinspace A_H>.80$

where

\hspace{35pt} $X_i = \textit{average speed of system i}$

\hspace{35pt} $A_i = \textit{accuracy of system i (proportion of submissions above quality threshold)}$

Since $X_C = 0$ by assumption, our hypothesis about $\hat{\rho}$  reduces to 

\hspace{35pt} $\hat{\rho}=\frac{X_{HC}}{X_H}>1$
\vspace{6pt}

For more details about the experiment design, see SI Appendix Secs. S3.3 and S3.4.

To see whether the hypothesis was confirmed, we consider two alternative ways of calculating $\hat{\rho}$ : (a) the ratio of means method, and (b) the regression method.

\subsubsection{Ratio of Means method} The most straightforward way to calculate $\hat{\rho}$ is to compute the ratio of the average speeds for the HC and H conditions shown in Table \ref{tab:empstatistics}(a): 

\vspace{6pt}
\hspace{35pt} $\hat{\rho}=\frac{X_{HC}}{X_H}=1.17$
\vspace{6pt}

We can compute a confidence interval for $\hat{\rho}$ following~\cite{bonett2020confidence}, which shows that the HC condition is significantly faster than the H condition (95\%CI: [1.04, 1.32]; see SI Appendix Sec. S3.5.1 for robustness checks). Table~\ref{tab:empstatistics}(a) shows that the accuracy conditions for both conditions are also satisfied ($A_{HC}  , A_H  > .80$). In other words, this simple test shows that synergy is present in these human-GPT-3 systems.

\begin{table}[!ht]
 \caption{Speed and quality averages per condition (see Figures S8-S10 for full distributions).}
 \label{tab:empstatistics}
 \centering
 \begin{tabular}{lccc}
   \toprule
	
   \multirow{3}{*}{Condition}      & Speed   & Accuracy & \# of \\
   
   & (tasks/min) & (\% of successful  & Observations \\
   &             &  subjects) &  \\

   \midrule
   
   \textbf{(a) Study 2} & & & \\
    Human Only (H) & 0.027& 84\% & 97 \\
 Human-Computer (HC) & 0.032 & 88\% & 97 \\
   \textbf{(b) Study 3} & & \\
   Human-Computer (HC') & 0.030 & 95\% & 96\\

   \bottomrule
 \end{tabular}
\end{table}

\subsubsection{Regression method}
It is also possible to estimate a ratio of means and its confidence interval while controlling for various other factors such as task, task order, and subject. To do this, the ratio $\hat{\rho}$ can be estimated using the following generalized mixed-effects linear regression:
$$y_{ij}=\beta_0 +\beta_1C_{ij}+\beta_2T_j+\beta_3O_{ij}+v_i+\epsilon_{ij}$$

 In this case, we define $y_{ij}$ as the logarithm of speed for the $i^{th}$ subject on the $j^{th}$ task. This has the benefit of making the regression equation consistent with a multiplicative model of performance that enables us to interpret the exponential of each coefficient as the multiplicative  increase in speed due to the different independent variables. It also has the benefit of normalizing the distribution of observed values. 
 
  For our purposes, the most important coefficient is $\beta_1 = log(\hat{\rho})$, the coefficient for the effect of the treatment condition (HC) relative to the control condition (H); $\beta_0$ is the fixed intercept; $\beta_2$ and $\beta_3$ are the fixed coefficients to control for the effect of the task and of the order in which this task occurred (first or second); $C_{ij}$, $T_j$, and $O_{ij} \in \{0,1\}$ are indicator variables respectively indicating which of the two conditions, which of the two tasks, and which of the two task orders occurred for the $i_{th}$ subject and $j{th}$ task; $v_i$ is the random coefficient for the $i_{th}$ subject; and $\epsilon_{ij}$ is a Gaussian error term. For more details about the regression specification, see SI Appendix Sec. S3.6.

\begin{figure}[ht]
\centering
\includegraphics[width=.45\textwidth]{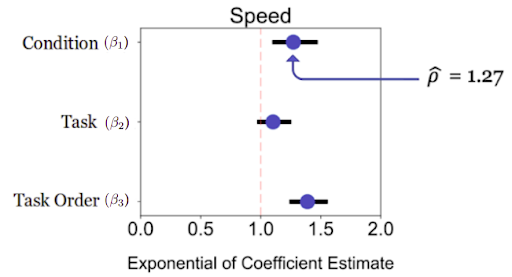}

\caption{Study 2 regression results.  Controlling for factors such as task and order, the humans and GPT-3 combination obtained statistically significant synergy as the performance improved by 27\% (a factor of 1.27 with 95\% CI of [1.10, 1.48]) relative to the performance of humans alone.}
\label{fig:regr}

\end{figure}

Our pre-registered hypothesis (\url{https://aspredicted.org/yx5m9.pdf}) was that the ratio ($\hat{\rho}$) obtained from the regression would be greater than 1.0 with 95\% confidence. As Figure \ref{fig:regr} shows, this hypothesis was confirmed with $\hat{\rho}$ = 1.27, CI[1.10, 1.48]. In other words, this regression version of the test also shows that the humans and GPT-3 combination obtained statistically significant synergy, improving by about a factor of 1.27 (a 27\% improvement) beyond the performance of humans alone. It is interesting to note that, in this case at least, controlling for the other factors such as task and order gives a higher ratio for $\hat{\rho}$ than calculating the simple ratio of means (see SI Appendix Section S3.7 for full regression results and Sections S3.5.2 and S3.5.3 for robustness checks).

\subsection{Study 3: Software generation by human non-programmers and GPT-3}

In the previous study, the subjects are experienced human programmers. But how might the results differ for humans without this skill? In Study 3, the subjects are humans who do not know how to code in HTML (“non-programmers,” see SI Section S3.3), and we have only one experimental condition. In this human-computer (HC) condition, the non-programmer subjects do the same two tasks using the same GPT-3 interface as the programmers in Study 2 (see SI Appendix Section S3.2). 

The order tasks are presented to subjects is randomized. As before, we assume that computers could not perform this task alone, and we also assume that the non-programmers (who do not know HTML) could not do it themselves alone either.

In other words, for Study 3, the conditions are:

\hspace{35pt}
\textit{H' = human non-programmers [assumed impossible, so not tested]}

\hspace{35pt}
\textit{C' = computers (assumed impossible, so not tested)}

\hspace{35pt}
\textit{HC' = human non-programmers with GPT-3 interface}

Our pre-registered hypothesis for this study was that the proportion of “successful” subjects would be greater than 50\% with 95\% confidence, where a subject is “successful” if they achieve at least 90\% of the total possible points (\url{https://aspredicted.org/yx5m9.pdf}).  Using the data in Table \ref{tab:empstatistics}(b), a one-sample proportion test (as specified in our pre-registration) shows that this hypothesis is strongly confirmed (z = 8.87, p~<<~.0001). 

As before, we also test the hypothesis that $\hat{\rho}> 1$, and we observe that

\vspace{6pt}
\hspace{35pt} $X_{HC'}=0.030 \textit{ tasks per minute}$ 

\hspace{35pt} $X_{H'}=X_{C'}=0 \textit{ (by assumption)}$
\vspace{6pt}

Since the denominator of $\hat{\rho}$ is 0, $\hat{\rho}$ will, of course, always be undefined, regardless of the value of $X_{HC'}$. However, in this context, it is reasonable to interpret $\hat{\rho}$ as an arbitrarily large number, which we denote here as “$\infty$”. In other words, we say that 

\hspace{35pt} $\hat{\rho}=\frac{X_{HC'}}{max(X_{H'},X_{C'})}=\frac{.030}{max(0,0)}=$ “$\infty"$
\vspace{6pt}

In practical terms, this means that non-programmer humans and the GPT-3 computer system  have very strong synergy. They can do something together that neither could do at all alone, and this is a very desirable result from the perspective of the $\hat{\rho}$ test. 

\subsubsection{Comparing programmers to non-programmers} Since there could be many factors affecting the differences between the populations of programmers and non-programmers in our study, we should be cautious about making causal inferences from a comparison of the results between these two populations. It is, however, interesting to calculate a value of $\rho$ by comparing the two here. 

As before, we do this in two ways. First, using the ratio of means method, we obtain

\vspace{6pt}
\hspace{35pt} $\rho=\frac{X_{HC}}{X_{HC'}}=1.07$
\vspace{6pt}

This suggests that the programmers are slightly faster than the non-programmers, but the confidence interval shows that this is not significantly different from 1 (95\% CI: [0.93,1.24]). 

Second, using the regression method, we obtain a similar value for $\rho$ of 1.015 (p = 0.902 and 95\% CI: [.81, 1.28]) and the results are also not significant (that is, $\rho$ is not significantly different from 1). In other words, it appears that not only does using GPT-3 allow the non-programmers in our study to do a task they could not otherwise have done, it allows them to do the task as fast as experienced programmers do. (see SI Appendix Sec. S3.7 for full regression results).

\subsubsection{Calculating $\rho$ for cost} We have focused so far on using $\rho$ to analyze differences in speed, but it is also possible to use it to analyze other performance dimensions. For example, one obvious question here involves cost: If non-programmers are now able to do a task using GPT-3 that previously would have required programmers, could it be more economical to use non-programmers for this task?
 
To obtain a suggestive answer to this question, we estimated the costs, $C_i$, of performing the task in condition $i$ by taking into account (a) the time each subject spent doing the task, (b) the cost of this time based on that subject’s individual hourly rates as shown on the UpWork.com site at the time of the experiment, (c) the number of calls to GPT-3 each subject made, and (d) the average cost of GPT-3 calls at the time of the experiment (see SI Appendix Sec. S3.8 for details). The resulting cost estimates are shown in Table \ref{tab:costs}.  

\begin{table}[!ht]
 \caption{The estimated costs for each condition. Hourly rate, time spent on each task, number of calls to GPT-3 and total cost to complete the task.}
 \label{tab:costs}
 \centering
\begin{tabular}{ l | c  c  c  c }
 \toprule
 
 \multirow{2}{*}{Condition}      & Rate/hr   & Mins/task & GPT-3 & Cost \\
   
   &  & & calls  &  \\

 \hline
 Non-programmers \& GPT-3 ($C_{HC'}$) & \$11.40 & 41.91 & 28.5 & \$8.82\\ 
 Programmers \& GPT-3 ($C_{HC}$) & \$16.28 & 38.00  & 24.0 & \$10.57\\  
 Programmers alone ($C_H$)& \$16.28 &  42.47 & - & \$10.92\\ 
  \bottomrule
\end{tabular}
\end{table}

Using the ratio of means method to estimate $\rho$ for the two conditions in Study 1, we obtain

\vspace{6pt}
 \hspace{35pt} $\rho=\frac{C_{HC}}{C_H}=0.97$
 \vspace{6pt}
 
 The result is not significantly different from 1 (95\% CI: [.87, 1.11]). The results of the regression method are similar:  $\rho$ = 0.890 (p = 0.109 and 95\% CI: [.77, 1.02]). So the use of GPT-3 does not appear to significantly reduce the total cost for programmers developing HTML code in this study.
 
 It is also interesting to calculate a different version of $\rho$ for the comparison between programmers with GPT-3 in Study 1 and non-programmers with GPT-3 in Study 2. Using the ratio of means method, we obtain
 
  \vspace{6pt}
  \hspace{35pt} $\rho=\frac{C_{HC}}{C_{HC'}}=1.20$
  \vspace{6pt}
  
  \noindent and this is not significantly different from 1 (95\% CI: [0.98, 1.47]). 
  
  However, when we estimate this version of $\rho$ using the regression method, we see that the cost for programmers was significantly greater than for non-programmers ($\rho$ = 1.40, p = .010, and 95\% CI: [1.09, 1.81]). In other words, the more sensitive regression test allows us to see that non-programmers using GPT-3 can create the HTML code for websites at significantly less cost than experienced programmers (see SI Appendix Sec. S3.9 for cost regressions). We’ll discuss some of the implications of this finding in the Discussion section.

\section{Discussion}

In this work, we proposed and demonstrated an approach to systematically studying the performance of human-computer systems by focusing on the ratio ($\rho$) of improvement that human-computer systems achieve relative to various baselines.

\subsection{Summarizing the state-of-the-art in human-computer groups}

In Study 1, we used this approach to analyze a comprehensive sample of recently published studies of human-computer systems and found that (a) the average ratio of improvement when humans and computers worked together versus alone was only about 1, which corresponds to no change at all, (b) the majority of experiments (62\%) did not observe any performance improvement, and (c) the maximum improvement was less than a factor of 1.36.

Of course, we analyzed only a limited sample of all the potentially relevant studies published over the recent decades, but our tentative explorations of other studies and other years suggest that these results are typical of the broader literature. 

These results raise interesting questions: Why aren’t the  ratios higher? Why don’t these studies find more synergy between people and computers? 

One possible answer is simply that many of the researchers in these studies may have been investigating other issues and not trying to maximize the improvements they achieved with human-computer systems. 

Another possible answer is that researchers may have intentionally avoided studying situations where finding a large improvement would have seemed obvious and uninteresting. It would probably seem obvious to computer scientists, for example, that a human programmer using high-level programming languages and standard compiler software can generate useful computer programs much faster than a programmer who had to write only machine language code. And a computer alone could not write the software at all because it wouldn't even know what to do without instructions from a human. So the combination of people and computers together can generate useful software at a level of performance that would have been impossible for either alone. But researchers might feel that showing this would be so unsurprising as to be uninteresting.  

There is also another possible reason why the results in Study 1 aren't more positive: Perhaps achieving substantial synergies among people and computers is harder than many people think. Perhaps it requires, not just new kinds of software, but also new ways of configuring groups that include people and computers. And perhaps it needs more systematic, focused attention from researchers than it has, so far, received. We hope that the approach we have proposed in this paper can help facilitate that focus.

\subsection{Experimenting with a massive, state-of-the-art AI system}
While the studies summarized in Study 1 included numerous relatively recently developed AI systems, none of them used the massive GPT-3 system which was only  released in mid-2020. Could its surprisingly general capabilities lead to higher values of $\rho$ than those summarized in Study 1?

In Study 2, we found that when we used this system to help human programmers generate software, the value of $\hat{\rho}$ was about 1.3, corresponding to a speed improvement of about 30\%. This  was a statistically significant level of what we define as \textit{synergy}, and it would have been one of the highest ratios in Study 1. (Even when calculated using the simpler ratio of means method, $\hat{\rho}$ would have been about 90th percentile in Study 1.) 

We also found that human \textit{non-}programmers using GPT-3 were able to do the same task as the programmers even though we assume that neither the human non-programmers nor today's AI algorithms could have done this task alone. In other words, this human-computer combination is an example of \textit{extreme synergy} ($\rho>>1$) where people and computers together can do something neither could do alone. 

Finally, we found that, in the samples of programmers and non-programmers we studied, the non-programmers using GPT-3 were able to create websites about as fast and at significantly less cost than the programmers. This would presumably be very desirable for managers and owners of organizations and for the newly-empowered non-programmers themselves. But it might also be seen as a form of deskilling for the programmers whose jobs could now be performed by people with less skill--and for lower compensation~\cite{attewell1987deskilling, downey2021partial}.

\subsection{A complement to the Turing test} As noted in the Introduction, generations of computer scientists have been inspired by the Turing test to try to create computers that can equal human performance. And to a significant degree, the vision of computers that can replicate--and potentially replace--humans has dominated much public discussion and business  decision-making in recent years. 

But as many observers since at least the 1960s have pointed out, computers can also be used to \textit{augment} human intelligence, not just to replace it [e.g.,~\cite{licklider1960man, engelbert1963conceptual, malone2018superminds, malone2020artificial}]. And, perhaps, our focus on the Turing test has even prevented us from recognizing and developing these possibilities in ways that would have been better for business and society~\cite{acemoglu2021, siddarth2021}. 

We believe that the test proposed here, by focusing on and quantitatively measuring the \textit{combined} performance of people and machines can help correct this imbalance and lead to better economic and societal uses of computers.

\textit{Contests for the best human-computer performance}. For example, one way this test could help stimulate progress is by having contests like the autonomous vehicle contests sponsored by DARPA that were so useful in the development of autonomous vehicles (e.g., \cite{rouff2011experience}). But in this case, the contests would not be about seeing which teams could create computers to equal human performance; they would be about creating human-computer systems to get the highest possible values of $\rho$ for various tasks.  

The contest organizers might, for example, identify the task to be done and specify (a) the pool of human participants, and (b) the specific hardware and software platforms that contestants could use. The goal of the contestants would then be to design human-computer systems that would produce the highest values for $\rho$.

Among other things, such contests would provide opportunities for constructive collaborations between computer scientists and social scientists that could be of very substantial practical value.  These multi-disciplinary teams would not only develop new algorithms and other technologies but also design and study new ways of configuring human-computer systems to perform various tasks more effectively.

\subsection{Measuring specialized collective intelligence} 
In addition to its use in analyzing the performance of human-computer systems, the $\rho$ test can also be thought of  as a measure of the collective intelligence of such systems. Even though some measures of collective intelligence focus on a group’s \textit{general} collective intelligence, that is, the group’s ability to perform a wide range of different tasks~\cite{woolley2010, Riedl2021}, it is also possible to measure a group’s \textit{specialized} collective intelligence, that is, its ability to perform a single, specific task~\cite{malone2018p24}. In this sense, we can view the $\rho$ test as a way of measuring the specialized collective intelligence of human-computer groups compared to relevant baselines. 

For instance, each of the experiments in Studies 1, 2, and 3 focuses on a specific task, such as developing software or answering open domain questions. In each of these cases, the $\hat{\rho}$ ratios we calculated can be viewed as measures of the specialized collective intelligence of human-computer groups performing these tasks compared to the specialized individual intelligences of humans and computers doing the same tasks.

More generally, we can also view the $\hat{\rho}$ measure as one simple way of measuring the benefits of having diverse types of intelligence in a group~\cite{page2008}. In other words, instead of just measuring the benefits of combining two types of intelligence (humans and computers) for given tasks, we can use a generalization of $\hat{\rho}$ to measure the benefits of combining other types of intelligent actors, such as men and women~\cite{woolley2010}, people with different cognitive styles~\cite{aggarwal2019}, or even people and animals~\cite{Darwiche2018}. To do this, we could generalize the numerator of $\hat{\rho}$ to be the performance of a diverse group (e.g., visual thinkers and abstract thinkers together) and the denominator to be the maximum performance of homogeneous groups (e.g., visual thinkers only and abstract thinkers only). Then, $\hat{\rho}$ would measure the benefits of the diverse group compared to the homogeneous groups. 

And in all these cases, the various $\rho$ measures have the benefits of intuitiveness and comparability described above.

\subsection{Conclusion} 
In summary, we hope that by scientifically measuring the performance of human-computer systems, the test we have proposed here will spur progress in developing such systems and in using them much more widely in our economies and societies. We also hope that such systems may someday be so much more effective than what we have today that we will call them ``superintelligent.'' 

\vspace{18pt}

{\large \textbf{Acknowledgments}}

This work was supported, in part, by Toyota Research Institute and the MIT Quest for Intelligence. The authors would also like to thank Dean Eckles for statistical advice and Raza Abbas for software development assistance.

\bibliographystyle{ieeetr}
\bibliography{main}  





\end{document}


\setcounter{page}{1}
\renewcommand{\thepage}{S\arabic{page}}
\renewcommand{\thetable}{S\arabic{table}}
\renewcommand{\thefigure}{S\arabic{figure}}

\section*{Supplementary Information}
\section*{S1 Approach}

\subsection*{S1.1. When is the ratio of means a useful measure of effect size?}

The basic requirement for a ratio of means to be appropriate is that the $X$'s be measured on a ratio scale \cite{Stevens1946} with a ``real 0'' that signifies the absence of whatever is being measured \cite{GrissomKim2012ch2}. For instance, a measure of 0 for length, weight, or speed signifies the absence of those qualities, and an object that is 2 meters long  is twice as long as one that is 1 meter long. But an object that has a temperature of $100^o$F on the Fahrenheit scale is not, in any useful sense, twice as hot as one that is $50^o$F. (This would be true, however, if the temperatures were measured on the Kelvin (K) scale which has an absolute 0.)

However, there are cases that are not strictly ratio scales within the Stevens taxonomy \cite{Stevens1946}, but where a ratio of means measurement can still be useful. For example, proportions (including percentages and probabilities) don't satisfy Stevens's strict criteria for a ratio scale  \cite{MostellerTukey1977, VellemanWilkinson1993}. One cannot, for instance, double a percentage without distorting its meaning. And when a student gets a 0 on a test, that doesn't necessarily mean that the student has no knowledge whatsoever of the subject to which the test pertains. But it is still often useful to analyze ratios of means for proportions, and we do so here.  

In fact, it may even be useful to analyze ordinal data (such as Likert scales) with a ratio of means. For instance, Stevens himself, who formulated the widely used taxonomy of scale types, said, ``As a matter of fact, most of the scales used widely and effectively by psychologists are ordinal scales. In the strictest propriety the ordinary statistics involving means and standard deviations ought not to be used with these scales... On the other hand, ... there can be invoked a kind of pragmatic sanction: in numerous instances it leads to fruitful results'' \cite{Stevens1951, VellemanWilkinson1993}. Examples of treating data from ordinal scales as ratio scales in this way include \cite{Buttonetal2015, Kline2005}, and we do so here.  

Another factor to consider in judging whether analyzing the ratio of means is appropriate in a given situation is whether the underlying process being measured can be usefully modeled as a multiplicative, rather than additive, process. For example, in Sec. S3.6 below, we provide a simple and very general model of performance as a multiplicative function of factors such as the ability of subjects and the difficulty of tasks.

\subsection*{S1.2. Metric transformations}

\subsubsection*{S1.2.1. Desirable lower bound}

The full version of the desirable lower bound transformation is 

\vspace{6pt}
\hspace{35pt} $f(X) = \frac{1}{X - X_{min}}$
\vspace{6pt}

\noindent where the lower bound of $X_{min}$ is usually 0. In cases where the lower bound is not 0, this transformation automatically rescales the original metric to have a lower bound of 0. And the transformed metric has a lower bound of 0 as X approaches $\infty$. 

As noted in the main text, this transformation is useful to make metrics where lower values are more desirable comparable to ones where higher values are better. It also has a side effect of emphasizing small improvements as the lower bound is approached (see Sec. S1.2.3 below).

\subsubsection*{S1.2.2. Desirable upper bound}

To derive the desirable upper bound transformation, we begin by rescaling X so that its transformed upper bound is 1:
     
\vspace{6pt}
\hspace{35pt} $X' = f(X) = \frac{X}{X_{max}}$
\vspace{6pt}

\noindent where $X_{max}$ is the original upper bound. Then we further transform the resulting value with an equation that is symmetrical to the one for the desirable lower bound transformation. In other words, we take the reciprocal of the distance to the bound:
     
\vspace{6pt}
\hspace{35pt} $f'(X') = \frac{1}{1 - X'}$
\vspace{6pt}.

But this transformation has the disadvantage that its minimum when X' = 0 is 1, which means that the transformed values would not have a "real 0." So to correct for this, we use a different transformation which subtracts 1:

\vspace{6pt}
\hspace{35pt} $f''(X') = \frac{1}{1 - X'} - 1$

\hspace{67pt} $= \frac{1}{1 - X'} - \frac{1 - X'}{1 - X'}$

\hspace{67pt} $= \frac{X'}{1 - X'}$.
\vspace{6pt}

Interestingly, this transformation is equivalent to the calculation of odds used in calculating an \textit{odds ratio}. That means that the ratio of means effect size measurement with this transformation is equivalent to an effect size measurement with an odds ratio \cite{GrissomKim2012ch8}.

\vspace{6pt}
\noindent \textit{Effect of the desirable upper bound transformation}. Another interesting feature of the desirable upper bound transformation is that it always makes the $\rho$ and $\hat{\rho}$ ratios more extreme, that is, further from 1. 

To see why analytically, first note that the transformation does not change which element in the denominator of $\hat{\rho}$ will be the maximum because the transformation does not change the inequality relationship between any two variables a and b.

To see this, let

\vspace{6pt}
\hspace{35pt}
$a' = \frac{a}{1 - a}$

\hspace{35pt}
$b' = \frac{b}{1 - b}$ 
\vspace{6pt}

Then, if $a > b$, the numerator of $a'$ is greater than the numerator of $b'$, and the denominator of $a'$ is less than the denominator of $b'$, so $a' > b'$. Similar reasoning applies for $a < b$, and $a = b$.

Now to see that the transformation increases the distance of a $\rho$ ratio from 1, let

\vspace{6pt}
\hspace{35pt}
$\rho = \frac{x}{y} = \textit{ratio of the untransformed values}$

\hspace{35pt}
$\rho' = \frac{\frac{x}{1-x}}{\frac{y}{1-y}} =  \textit{ratio of the transformed values}$

\vspace{6pt}
\hspace{35pt}
$R = \frac{\rho'}{\rho} = \frac{\frac{1}{1-x}}{\frac{1}{1-y}} = \frac{1-y}{1-x}$
\vspace{6pt}

Thus, if x > y, then R > 1, and the ratio of the transformed values is greater than the ratio of the untransformed ones (i.e., further above 1). The opposite is true when x < y. And if x = y, then R = 1, and the ratio of the transformed values equals the ratio of the untransformed ones.

\subsubsection*{S1.2.3. Emphasizing small improvements near a bound}

A potential benefit of both the upper and lower bound transformations is that they \textit{emphasize small improvements as a metric approaches its desirable limit}. This can be especially useful when approaching a limit is both increasingly difficult and increasingly desirable.

For example, if $X$ represents the proportion of lives saved by a medical treatment, the fraction of airplane accidents prevented, or the number of costly failures in critical infrastructures (such as computer servers), then it can be both extremely difficult and extremely valuable to approach the limits in these situations.

In other situations, however, small changes near a limit may have no particular value. For instance, in producing a product with limited demand (such as a customized report), the benefit of reducing the cost from \$0.01 to \$0.001 may be irrelevant even though it produces a factor of 10 improvement according to the transformed metric.

\section*{S2 Study 1}

\subsection*{S2.1. Systematic review and analysis}

\subsubsection*{S2.1.1. Search strategy}

To select the studies for our analysis, we followed the procedures of a systematic review~\cite{Kitchenham2004ProceduresFP, Liberatib2700}.  As such, we began by developing the search string and distilling the facets of studies that evaluated the performance of a human-computer system.  We required the following: (1) a human component, (2) a computer component, (3) a collaboration component, and (4) an experiment component.  Given the multidisciplinary nature of our research question, papers published in different venues tended to refer to these components under various names~\cite{Dellermann2019TheFO,Gerber2020ConceptualizationOT} To ensure comprehensive coverage, we compiled a list of synonyms and abbreviations for each component and then combined these terms with Boolean operations, resulting in the following search string:

\begin{enumerate}
    \item Human agent:  “human” OR “expert” OR “participants”
    \item AI agent:  “AI” OR “artificial intelligence” OR “ML”, or “machine learning”
    \item Collaboration:  “collaboration” OR “teamwork” OR “combination” OR “complement” OR “synergy” OR “assist” OR “aid” OR “hybrid” OR “loop”
    \item Experiment:  “experiment” OR “user study” OR “user evaluation” OR “empirical study”
\end{enumerate}

Search term: (1) AND (2) AND (3) AND (4)
\vspace{6pt}

We performed this search in the Association for Computing Machinery Digital Library (ACM DL), Institute of Electrical and Electronics Engineers Xplore Digital Library (IEEE Xplore), Association for Information Systems eLibrary (AISeL), Web of Science, and arXiv to encompass the major computer science, engineering, information systems, and social science conferences and journals.  To focus on current forms of artificial intelligence, we limited the search to studies published in 2021.

\subsubsection*{S2.1.2. Selection criteria}

Next, we applied the following criteria to select studies that fit our research questions. Specifically, the paper needed to include an original experiment that evaluated some instance in which a computer helped a human perform a task, and it needed to report  the performance of (1) the human alone, (2) the computer alone, and (3) the human-computer team according to some quantitative measure(s).  As such, we excluded meta and literature reviews, commentaries, opinions, and simulations.  Furthermore, we required the study to be written in English.  

Through these selection criteria, we identified 20 relevant papers from the initial search.  Among this set, we performed a forward and backward search, which found 5 new studies to include in our final review.  Notably, most of these papers reported the results of multiple experiments, so our final list included 25 unique papers and 79 unique experimental measurements. 

\subsubsection*{S2.1.3. Data extraction}

To address our research questions, we extracted the performance measure(s) for each experiment in each study and recorded the performance of the human alone ($X_{H}$), the computer alone ($X_{C}$), and the human-computer team ($X_{HC}$) according to the measure(s) reported in the paper.  Note that if the authors assessed performance in a single experiment according to multiple metrics (e.g., specificity and sensitivity), then we included all the different metrics as separate entries in our data.  For each metric, we also calculated $\hat{\rho}$, which included the desirable lower bound transformation wherever applicable, and $\hat{\rho}'$, which included both the desirable lower bound and upper bound transformations wherever they were applicable.  Table S1 summarizes our results.

\begin{small}
\begin{longtable}{ccccccccc}
\toprule
\textbf{Study} &\textbf{Task} &\textbf{Measure} &\textbf{$X_{H}$} &\textbf{$X_{C}$} &\textbf{$X_{HC}$} &\textbf{$\hat{\rho}$} &\textbf{$\hat{\rho}'$} \\\midrule
\cite{Gonzalez2021DoEH} &Answer questions &Accuracy &57\% &50\% &78\% &1.36 &2.59 \\
\cite{Gonzalez2021DoEH} &Answer questions &Accuracy &57\% &50\% &76\% &1.32 &2.32 \\
\cite{Gonzalez2021DoEH} &Answer questions &Accuracy &57\% &50\% &75\% &1.31 &2.21 \\
\cite{Gonzalez2021DoEH} &Answer questions &Accuracy &57\% &50\% &71\% &1.25 &1.86 \\
\cite{Gonzalez2021DoEH} &Answer questions &Accuracy &57\% &50\% &70\% &1.23 &1.78 \\
\cite{Gonzalez2021DoEH} &Answer questions &Accuracy &57\% &50\% &68\% &1.19 &1.60 \\
\cite{Holstein2021DesigningFH} &Teach students math concept &Accuracy &35\% &51\% &60\% &1.18 &1.44 \\
\cite{Fang2021DesignAA} &Detect objects in an image &Recall &77\% &64\% &89\% &1.17 &2.55 \\
\cite{Leimkhler2021DeepLO} &Label visible car scratches &Recall &51\% &43\% &59\% &1.16 &1.40 \\
\cite{Leimkhler2021DeepLO} &Label visible car scratches &Precision &53\% &53\% &61\% &1.15 &1.40 \\
\cite{Fang2021DesignAA} &Detect mulitple objects in an image &Recall &77\% &64\% &86\% &1.12 &1.82 \\
\cite{groh2022deepfake} &Detect deepfakes &Accuracy &66\% &65\% &73\% &1.11 &1.39 \\
\cite{Desmond2021IncreasingTS} &Label data &Accuracy &72\% &44\% &79\% &1.10 &1.46 \\
\cite{Desmond2021IncreasingTS} &Label data &Accuracy &72\% &59\% &78\% &1.08 &1.38 \\
\cite{Bondi2021RoleOH} &Classify conservation images &Accuracy &58\% &50\% &62\% &1.06 &1.16 \\
\cite{Fang2021DesignAA} &Detect mulitple objects in an image &Recall &77\% &64\% &80\% &1.05 &1.24 \\
\cite{Fgener2021WillHB} &Classify images &Accuracy &68\% &77\% &80\% &1.04 &1.20 \\
\cite{Fgener2021WillHB} &Classify images &Accuracy &68\% &77\% &80\% &1.04 &1.19 \\
\cite{Bondi2021RoleOH} &Classify conservation images &Accuracy &58\% &50\% &60\% &1.03 &1.09 \\
\cite{Shin2021AIassistanceFP} &Classify endoscopic images &Sensitivity &93\% &89\% &96\% &1.03 &1.75 \\
\cite{Shin2021AIassistanceFP} &Classify endoscopic images &Sensitivity &93\% &89\% &95\% &1.02 &1.40 \\
\cite{Baudel2020AddressingCB} &Decide if Titanic passenger survived &Accuracy &72\% &75\% &77\% &1.02 &1.09 \\
\cite{Baudel2020AddressingCB} &Decide if Titanic passenger survived &Accuracy &72\% &75\% &77\% &1.02 &1.09 \\
\cite{Baudel2020AddressingCB} &Decide if Titanic passenger survived &Accuracy &72\% &75\% &76\% &1.02 &1.07 \\
\cite{Baudel2020AddressingCB} &Play kitchen game &Steps &37.82 &34 &38.37 &1.01 &0.89 \\
\cite{Baudel2020AddressingCB} &Decide if Titanic passenger survived &Accuracy &72\% &75\% &76\% &1.01 &1.06 \\
\cite{Pereira2021TowardsAH} &Classify programming problems &F1 &54\% &86\% &87\% &1.01 &1.09 \\
\cite{Gao2021EvaluatingHH} &Answer conversational questions &Quality &3.74 &3.68 &3.77 &1.01 &1.03 \\
\cite{Bastani2021ImprovingHD} &Play kitchen game &Steps &23.86 &20 &24.04 &1.01 &0.83 \\
\cite{Fgener2021WillHB} &Classify images &Accuracy &72\% &77\% &78\% &1.01 &1.03 \\
\cite{Liu2021UnderstandingTE} &Predict violation of pretrial release terms &Accuracy &56\% &55\% &56\% &1.01 &1.01 \\
\cite{Gao2021EvaluatingHH} &Answer conversational questions &Quality &3.74 &3.87 &3.89 &1.01 &1.02 \\
\cite{Liu2021UnderstandingTE} &Predict recidivism &Accuracy &55\% &56\% &56\% &1.00 &1.01 \\
\cite{Liu2021UnderstandingTE} &Predict recidivism &Accuracy &55\% &56\% &56\% &1.00 &1.00 \\
\cite{Zeleznik2021DeeplearningST} &Segment heart images &Dice &0.92 &0.92 &0.92 &1.00 &1.00 \\
\cite{Fgener2021WillHB} &Classify images &Accuracy &72\% &77\% &77\% &1.00 &0.98 \\
\cite{Liu2021UnderstandingTE} &Predict violation of pretrial release terms &Accuracy &56\% &55\% &56\% &0.99 &0.99 \\
\cite{Fang2021DesignAA} &Detect mulitple objects in an image &Precision &98\% &86\% &97\% &0.99 &0.78 \\
\cite{Fang2021DesignAA} &Detect mulitple objects in an image &Precision &98\% &86\% &97\% &0.99 &0.70 \\
\cite{Bondi2021RoleOH} &Classify conservation images &Accuracy &58\% &50\% &58\% &0.99 &0.98 \\
\cite{Pavoni2022TagLabAA} &Annotate image &Accuracy &95\% &88\% &94\% &0.99 &0.79 \\
\cite{Kleinberg2021HowHI} &Judge truthful and deceptive statements &AUC &52\% &75\% &74\% &0.99 &0.95 \\
\cite{Pavoni2022TagLabAA} &Annotate image &Accuracy &95\% &88\% &94\% &0.99 &0.76 \\
\cite{Bastani2021ImprovingHD} &Play kitchen game &Steps &37.82 &34 &37.14 &0.98 &0.92 \\
\cite{Fang2021DesignAA} &Detect mulitple objects in an image &Precision &98\% &86\% &96\% &0.98 &0.54 \\
\cite{Shin2021AIassistanceFP} &Classify endoscopic images &Specificity &85\% &91\% &89\% &0.98 &0.82 \\
\cite{Kleinberg2021HowHI} &Judge truthful and deceptive statements &Accuracy &50\% &69\% &67\% &0.97 &0.91 \\
\cite{Levy2021AssessingTI} &Annotate clinical text &Recall &76\% &83\% &80\% &0.96 &0.82 \\
\cite{Bastani2021ImprovingHD} &Play kitchen game &Steps &23.86 &20 &22.99 &0.96 &0.87 \\
\cite{Fgener2021WillHB} &Classify images &Accuracy &72\% &77\% &74\% &0.96 &0.85 \\
\cite{Liu2021UnderstandingTE} &Predict person's profession &Accuracy &68\% &77\% &74\% &0.96 &0.85 \\
\cite{Liu2021UnderstandingTE} &Predict person's profession &Accuracy &68\% &77\% &73\% &0.95 &0.82 \\
\cite{Jesus2021HowCI} &Predict risk of fraud payment transactions &Accuracy &62\% &57\% &59\% &0.95 &0.87 \\
\cite{Skirzynski2021AutomaticDO} &Mouselab-MDP &Score &28.41 &39.692 &36.8 &0.93 &0.09 \\
\cite{Gao2021EvaluatingHH} &Answer conversational questions &Quality &3.74 &3.1 &3.45 &0.92 &0.75 \\
\cite{Levy2021AssessingTI} &Annotate clinical text &Recall &76\% &83\% &76\% &0.92 &0.65 \\
\cite{Jesus2021HowCI} &Predict risk of fraud payment transactions &Accuracy &62\% &57\% &57\% &0.91 &0.80 \\
\cite{Liu2021UnderstandingTE} &Predict recidivism &Accuracy &60\% &66\% &59\% &0.91 &0.77 \\
\cite{Liu2021UnderstandingTE} &Predict violation of pretrial release terms &Accuracy &61\% &68\% &62\% &0.90 &0.74 \\
\cite{Gao2021EvaluatingHH} &Answer conversational questions &Quality &3.74 &2.84 &3.35 &0.90 &0.68 \\
\cite{Liu2021UnderstandingTE} &Predict recidivism &Accuracy &60\% &66\% &59\% &0.89 &0.75 \\
\cite{Shin2021AIassistanceFP} &Classify endoscopic images &Specificity &65\% &91\% &81\% &0.89 &0.47 \\
\cite{Liu2021UnderstandingTE} &Predict violation of pretrial release terms &Accuracy &61\% &68\% &61\% &0.89 &0.71 \\
\cite{Jain2021DevelopmentAA} &Diagnose dermatologic conditions &Accuracy &46\% &66\% &58\% &0.88 &0.71 \\
\cite{Kyriacou2021LearningTW} &Classify recyclables &Accuracy &86 &94.50 &82.5 &0.87 &1.00 \\
\cite{Liu2021UnderstandingTE} &Predict person's profession &Accuracy &64\% &84\% &73\% &0.87 &0.52 \\
\cite{Liu2021UnderstandingTE} &Predict person's profession &Accuracy &64\% &84\% &72\% &0.86 &0.50 \\
\cite{Skirzynski2021AutomaticDO} &Mouselab-MDP &Score &4.17 &6.97 &5.95 &0.85 &0.59 \\
\cite{Skirzynski2021AutomaticDO} &Mouselab-MDP &Score &22.85 &29.77 &24.54 &0.82 &0.03 \\
\cite{Jesus2021HowCI} &Predict risk of fraud payment transactions &Accuracy &62\% &57\% &51\% &0.82 &0.64 \\
\cite{Kyriacou2021LearningTW} &Classify recyclables &Accuracy &86 &55.00 &68 &0.79 &1.00 \\
\cite{Kleinberg2021HowHI} &Judge truthful and deceptive statements &Accuracy &50\% &69\% &51\% &0.74 &0.47 \\
\cite{Kleinberg2021HowHI} &Judge truthful and deceptive statements &AUC &52\% &75\% &49\% &0.65 &0.32 \\
\cite{Jacobs2021HowMR} &Prescribe drug &Accuracy &36\% &67\% &37\% &0.55 &0.29 \\
\cite{Jacobs2021HowMR} &Prescribe drug &Accuracy &36\% &67\% &36\% &0.54 &0.28 \\
\cite{Jacobs2021HowMR} &Prescribe drug &Accuracy &36\% &67\% &35\% &0.53 &0.27 \\
\cite{Jacobs2021HowMR} &Prescribe drug &Accuracy &36\% &67\% &34\% &0.51 &0.25 \\
\cite{Buccinca2021ToTO} &Edit food ingredients &Accuracy &17\% &75\% &35\% &0.47 &0.18 \\
\cite{Buccinca2021ToTO} &Edit food ingredients &Accuracy &17\% &75\% &33\% &0.44 &0.16 \\
\bottomrule
\caption{Summary of studies in review.}
\label{tab:summary}
\end{longtable}

\begin{figure}[ht]
\centering
\includegraphics[width=.75\textwidth]{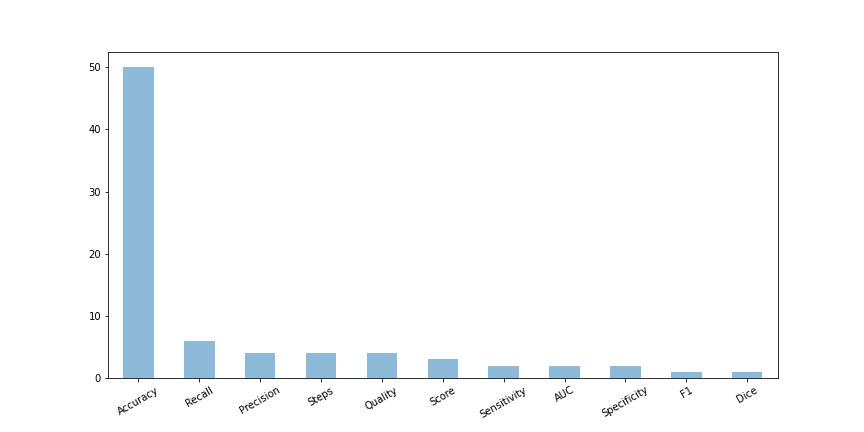}
\caption{The distribution of performance measures for the experimental results in our review ($n=79$).  Most studies evaluate the performance of human-computer teams according to accuracy, but we find a total of 11 different metrics used across all the studies.}
\label{fig:measures}
\end{figure} 

\subsubsection*{S2.1.4. Is this a meta-analysis?}

The goal of a standard meta-analysis is to estimate the “true” value of an underlying parameter (such as the beneficial effect of a particular drug) by calculating a weighted average of different measurements of that parameter from different studies~\cite{cochranehdbk}. In that sense, our study is not a standard meta-analysis because we did not attempt to construct a weighted average of the results from different studies. Instead, we did a very simple kind of meta-analysis in which we summarized the ratios from all the different studies by displaying their distribution and by reporting their median and (unweighted) mean. 

We did not attempt to compute weightings of the different studies because the studies we analyzed covered many different combinations of tasks, subjects, and technologies, and each combination could be expected to have a different “true” value for $\rho$. Thus any attempt to estimate a single underlying parameter for them all would be roughly analogous to trying to estimate the average weight of all the animals on earth–from tiny micro-organisms to large mammals. Even if one attempted to do this, it’s not clear whether one should weight different measurements based on (a) the statistical uncertainties in each measurement (as is common in standard meta-analyses), (b) the estimated number of animals of each type on earth, or (c) some other weightings. Furthermore, even if it were appropriate to estimate a single underlying value using the statistical techniques common in meta-analyses, most of the studies we analyzed did not include enough data to calculate such weighting statistics. 

We do still believe, however, that this simple way of displaying and summarizing the  ratios from recent studies provides novel and informative insights that set the stage for many kinds of future research.

\subsubsection*{S2.1.5. Robustness checks}
\textit{Effect of including multiple experimental measurements per paper}. In our main results, we include all experimental results from every study in our analysis.  Here, we also perform the analysis on only one experimental result per study, selecting the result that leads to the highest ratio (Figure \ref{fig:ratios_hist_both}).  By construction, among this subset of experiments, a larger proportion report findings of human-computer synergy according to the definition of a performance ratio greater than one (56\% versus 38\%), and the average and median ratios are slightly higher (0.98 versus 0.96 and 1.01 versus 0.99, respectively).  As such, our main findings still hold – a surprisingly small number of studies find evidence of human-computer synergy, and those that do find surprisingly small performance improvements.

\begin{figure}[ht]
\begin{center}
\includegraphics[width=.75\textwidth]{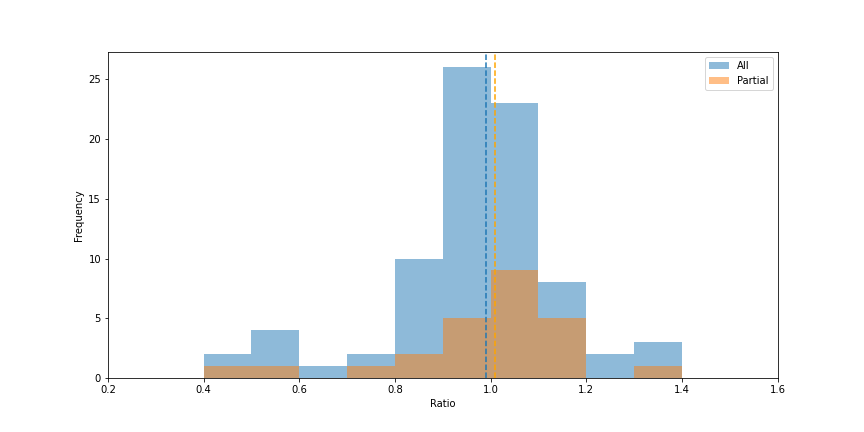}
\caption{The distribution of performance ratio values from all experimental results ($n=79$) and the top result per study ($n=25$).  The dotted lines correspond to the mean of the ratios.}
\label{fig:ratios_hist_both}
\end{center}
\end{figure} 
\end{small}

\textit{Effect of metric transformations}. To evaluate the effect of the metric transformations, we separately looked at the distribution of ratios for measures in which smaller values indicate better performance (desirable lower bound) and the distribution of ratios for measures in which larger values indicate better performance (desirable upper bound).  

\begin{figure}[ht]
    \centering
    \begin{subfigure}[b]{0.45\textwidth}
        \centering
        \includegraphics[width=\textwidth]{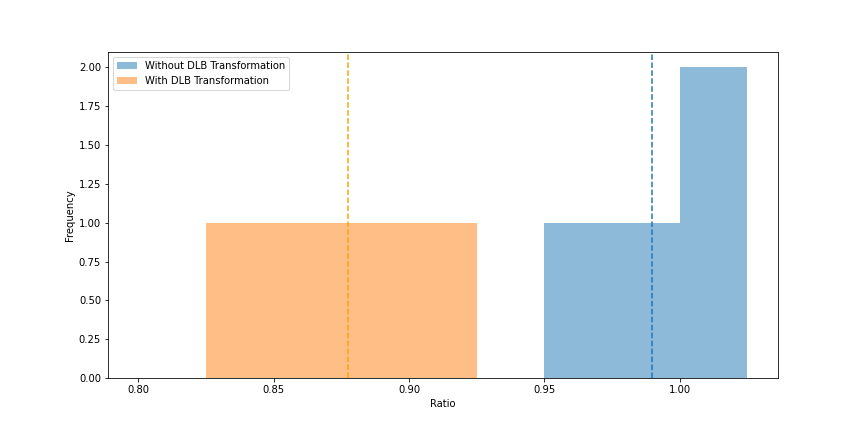}
        \caption{The distribution of ratios from experimental results with desirable lower bounds ($n=4$).}
    \end{subfigure}
        \begin{subfigure}[b]{0.45\textwidth}
        \centering
        \includegraphics[width=\textwidth]{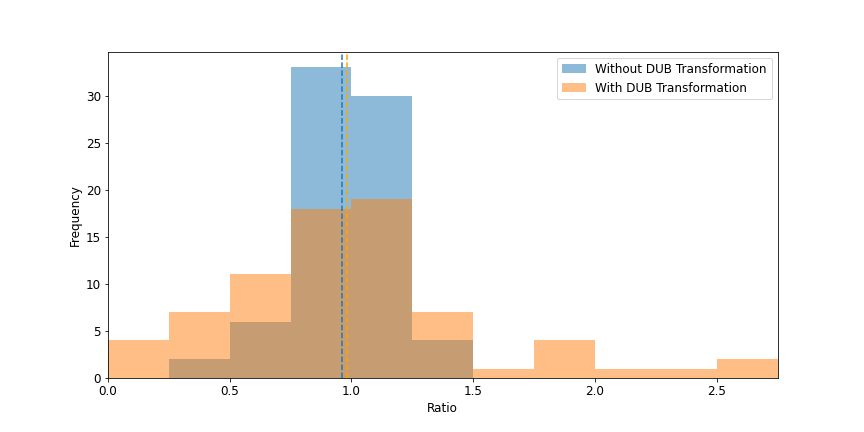}
        \caption{The distribution of performance ratios from experimental results with desirable upper bounds ($n=75$).}
    \end{subfigure}
\caption{The distribution of performance ratios from experimental results with and without transformations.  The dotted lines correspond to the mean of the ratios.}
\label{fig:ratios_hist_dup}
\end{figure}

Only four experimental results in our review involve a metric with a desirable lower bound, namely the number of steps required to complete a task (see Figure \ref{fig:ratios_hist_dup}a).  In this case, the transformation serves three functions: (1) it inverts the measure such that larger values correspond to better performance, (2) it changes which of the baselines in the denominator (human only or computer only) is selected for comparison, and (3) it rescales the measure so that small differences near the lower bound of 0 have a larger effect. Since the computer outperforms the human-computer team in all these experiments, the transformed ratios all fall below one, and the changes in magnitude from the untransformed values are relatively small.

The remaining experimental results all pertain to metrics with desirable upper bounds. As noted in Sec. S1.2, the effect of this transformation is to rescale the ratios to be further from 1, and this effect is visible in Figure \ref{fig:ratios_hist_dup}b.  As such, among this set, the percentage of experiments that exhibit human-computer synergy as defined by a ratio greater than one remains unchanged.  While the transformation widens the distribution of rho values, the median stays relatively the same (0.92 with versus 0.96 without the transformations) as does the mean (0.98 with versus 0.99 without the transformations).

\subsection*{S2.2. Survey}

To contextualize the findings from our review, we conducted an online survey with researchers actively pursuing empirical studies of human-computer teams.  In this section, we detail the survey design and evaluation, which follow the guidelines presented in~\cite{KitchenhamSurvey, KitchenhamSurveyTwo}.  

\subsubsection*{S2.2.1. Participant selection}

To target researchers with deep knowledge of human-computer teams, we sent the survey to the authors of the papers included in our review as well as authors from 2 additional papers that were later removed from our analysis because they did not include performance data for computers alone.  Since this set only includes studies published in the past year (2021), we expect these people to possess knowledge of the current state of the field.  Note that other studies such as~\cite{Diebold2014PractitionersAR} and~\cite{Jarvis2015WhatIC} perform similar strategies to elicit expectations from practitioners and researchers.  We sent emails to each author ($n=133$), but 11 addresses were no longer active, so we only delivered 122 invitations to complete the survey.  Additionally, we sent a friendly reminder about the survey after one week and two weeks.  After three weeks, we closed the survey.  We incentivized participation with a raffle for a \$200 Amazon gift card.

\subsubsection*{S2.2.2. Survey design}

\begin{figure}[ht]
    \centering
    \includegraphics[width=.5\textwidth]{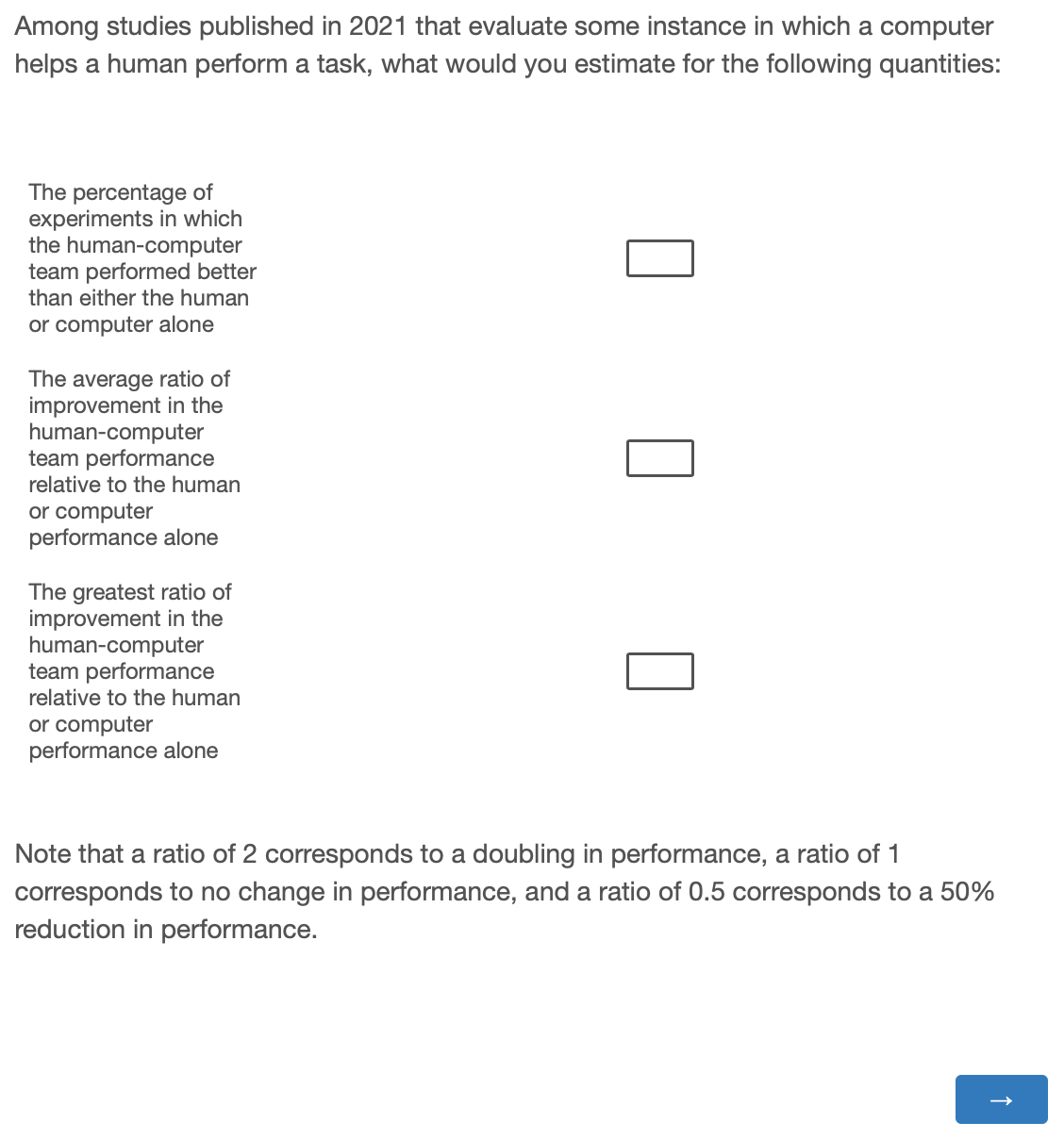}
    \caption{Screenshot of the survey page with questions for the participants.}
\label{fig:survey_q}
\end{figure}

The first block of the survey presented a consent form with information about the context of the study, expected completion time, and related matters.  In the next block, subjects answered the questions shown in Figure \ref{fig:survey_q}.  We asked for their responses via text boxes to mitigate potential anchoring effects~\cite{Gehlbach2012AnchoringAA}, and we only allowed them to submit numeric answers.  The final block of questions collected demographic information such as level of education, occupation, and general research discipline.  We made all questions optional, and we implemented the survey using Qualtrics. 

\pagebreak

\subsubsection*{S2.2.3. Participant demographics}

\begin{figure}[ht]
\centering
\begin{subfigure}[t]{.3\textwidth}
    \centering
    \includegraphics[width=\linewidth]{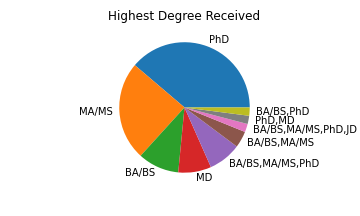}
    \caption{Participant highest level of education.}
\end{subfigure}
\hfill
\begin{subfigure}[t]{.3\textwidth}
    \centering
    \includegraphics[width=\linewidth]{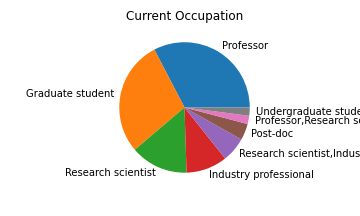}
    \caption{Participant occupation.}
\end{subfigure}
\hfill
\begin{subfigure}[t]{.3\textwidth}
    \centering
    \includegraphics[width=\linewidth]{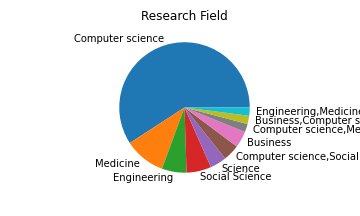}
    \caption{Participant research area.}
\end{subfigure}
\caption{Participant demographic information.}
\label{fig:survey_dem}
\end{figure}

A total of 49 people completed the survey for a response rate of 40\%.  Figure \ref{fig:survey_dem} provides an overview of the demographics of the participants.  Most respondents were currently pursuing or had already received a graduate degree, and most specialized in computer science.

\subsubsection*{S2.2.4. Survey results} 
\begin{figure}[ht]
\centering
\begin{subfigure}[t]{.3\textwidth}
    \centering
    \includegraphics[width=\linewidth]{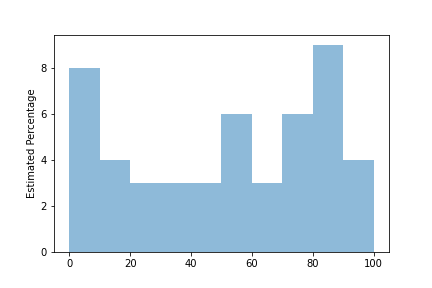}
    \caption{The percentage of experiments in which the human-computer team performed better than either the human or computer alone.}
\end{subfigure}
\hfill
\begin{subfigure}[t]{.3\textwidth}
    \centering
    \includegraphics[width=\linewidth]{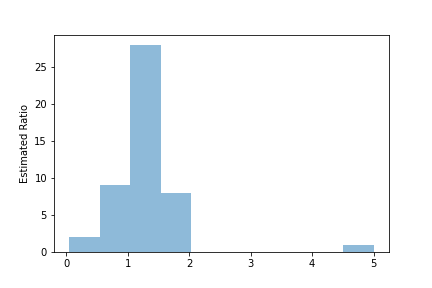}
    \caption{The average ratio of improvement in the human-computer team performance relative to the human or computer performance alone.}
\end{subfigure}
\hfill
\begin{subfigure}[t]{.3\textwidth}
    \centering
    \includegraphics[width=\linewidth]{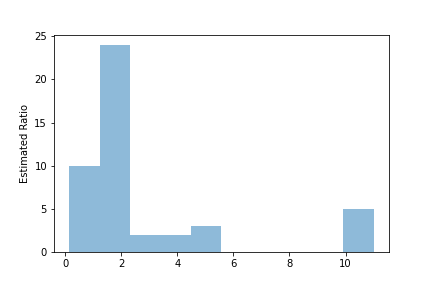}
    \caption{The greatest ratio of improvement in the human-computer team performance relative to the human or computer performance alone.}
\end{subfigure}
\caption{The distribution of participant responses to the questions in our survey.  Note that for readability purposes, (b) does not display one participant’s estimate of 20, and (c) does not display two participants’ estimates of 50 and 100.}
\label{fig:survey_results}
\end{figure}

Figure \ref{fig:survey_results} illustrates the distribution of the responses to the questions in the survey.  Given the number of extreme outliers in the responses, we highlight the median and interquartile ranges (IQRs) as opposed to the mean and standard deviations of the estimates for each question~\cite{RODRIGUES2017619}.  When asked to consider studies published in 2021 that evaluate some instance in which a computer helps a human perform a task, the median estimate of the percentage of experiments in which the human-computer team performed better than either the human or computer alone was 50\% (20 - 80\% IQR).  The median estimate of the average ratio of improvement was 1.2x (1.1 - 1.5x IQR), and the median estimate of the maximum ratio of improvement was 2.0x (1.5 - 4.0x IQR). 

\section*{S3 Studies 2 and 3}

\subsection*{S3.1. Web pages}

\begin{figure}[ht]
\centering
\begin{subfigure}[b]{.45\textwidth}
    \centering
    \includegraphics[trim=0 30 0 30,clip,width=\linewidth]{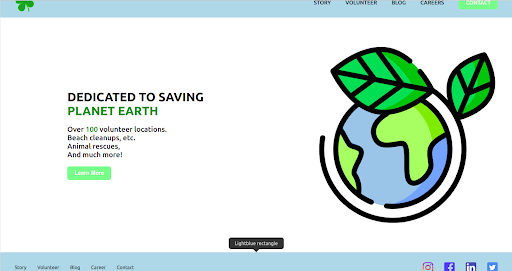}
    \caption{Saving Planet Earth, with 27 items.}
\end{subfigure}
\hfill
\begin{subfigure}[b]{.45\textwidth}
    \centering
    \includegraphics[trim=0 30 0 30,clip,width=\linewidth]{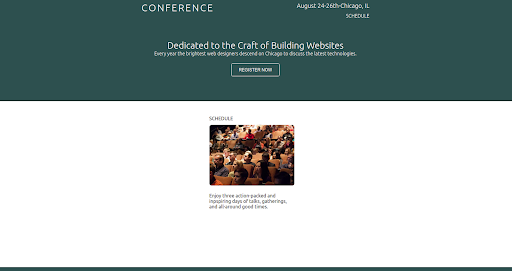}
    \caption{Conference Website, with 13 items.}
\end{subfigure}

\caption{Webpages used for experiment.}
\label{fig:mockups}
\vspace{-.4cm}
\end{figure}

To represent the space of possible web pages, we used two sample pages that varied in difficulty:

\begin{enumerate}
    \item "Saving Planet Earth" (Fig. \ref{fig:mockups}a) was designed by us and contains 27 components used in scoring the correctness of the code generated.
    \item "Conference Website" (Fig. \ref{fig:mockups}b) was adapted and simplified from a task given towards the end of an HTML course called Learn to Code (\url{https://learn.shayhowe.com/ practice/adding-media/index.html}) and contains only 13 components.
\end{enumerate}

In addition to the visual images shown in the figures, each image also included pop-up messages to specify the details of active elements such as buttons and links. Subjects could see these messages by rolling their mouse over the images. 

\subsection*{S3.2. GPT-3}

\subsubsection*{S3.2.1. GPT-3 description} GPT-3 stands for “Generative Pre-trained Transformer 3” \cite{gpt3}. It is a massive AI system that uses autoregressive machine learning techniques to generate many kinds of text. With over 175 billion parameters, it was extensively trained on a vast corpus of text including all of Wikipedia, many books, and much more material from the Internet.  It has outperformed existing state-of-the-art models in many benchmarks. Particularly impressive and relevant is its performance in the few-shot domain, where the model is conditioned to a specific task by providing only a few examples.

\subsubsection*{S3.2.2. Use of GPT-3 in these studies} We developed special software for our subjects to use as an interface to GPT-3.  This software uses the OpenAI API to interface with GPT-3, and it provides two primary additional features:

\begin{enumerate}
    \item \textit{Parameters and prompts.} To perform specific tasks, GPT-3 needs some detailed parameters and (usually) some examples to indicate what kinds of text outputs the system should generate in response to various kinds of inputs. See Sec. S3.2.3 for details of the parameters and examples we provided.
     \item \textit{User interface.} The user interface for GPT-3 that was used by subjects in both studies is shown in Figure 3. It works as follows: First, the human subject describes in English text a component needed in the web page (e.g., “Green button that says ‘Learn More’” as shown in the upper left corner of Figure 3). Then the GPT-3 system automatically generates the HTML code needed to produce that component (e.g., the code shown in the lower left corner of Figure 3). And, simultaneously, the interface we created adds that component to the replica of the web page shown on the right side of the screen (e.g., the green button shown in the middle of Figure 3). If the human thinks the displayed component is correct, the human can then move that component to its proper location in the replica of the web page.  If not, the human can (a) delete the component and try again to describe in English what the component should be, or (b) modify the HTML code directly (if the human knows enough to do so).

\end{enumerate}

\subsubsection*{S3.2.3. GPT-3 Parameters and Prompts}
For all our experiments we used the Open-AI API with the Davinci engine, a temperature of 0.3 and a max\_token length of 512.

\vspace{6pt}
\noindent \textit{Conditioning} GPT-3 is conditioned by prompting where each prompt includes a natural language description that the user would type followed by the HTML that would be returned by GPT-3:

\begin{itemize}
\item ``Hello world button''\\
<button style="padding: 12px 24px; border: 1px solid lightgray; background-color:
whitesmoke; border-radius: 4px; font-size: 14px; color: black; font-weight:
normal;">Hello World</button>

\item ``pink button that says Banana with white text color''\\
<button style="padding: 12px 24px; border: 1px solid lightgray; background-color: pink;
border-radius: 4px; font-size: 12px; color: white; font-weight:
normal;">Banana</button>

\item ``Large Google Pixel button with no background''\\
<button style="padding: 12px 24px; border: 1px solid lightgray; background-color:
transparent; border-radius: 4px; font-size: 24px; color: black; font-weight:
normal;">Google Pixel</button>

\item ``red square''\\
<div style="width: 100px; height: 100px; background-color: red;"></div>

\item ``orange rectangle''\\
<div style="width: 200px; height: 100px; background-color: orange;"></div>

\item ``paragraph that says I like apples''\\
 <p style="color: black; font-size: 14px;">I like apples</p>
 
 \item ``large text that says Black Circle in blue''\\
 <span style="color: blue; font-size: 24px;">Black Circle</span>
 
 \item ``Air-Conditioner in brown color''\\
 <span style="color: brown;">Air-Conditioner</span>
 
 \item ``My birthday is coming up soon!''\\
  <p>My birthday is coming up soon!</p>
  
  \item ``text that says Google and links to https://www.google.com/''\\
  <a href="https://www.google.com/" target="\_blank" style="color: black; text-decoration:
none;">Google</a>

\item ``FACEBOOK and links to https://www.facebook.com/''\\
<a href="https://www.facebook.com/" target="\_blank" style="color: black;
text-decoration: none;">FACEBOOK</a>
 
\item ``image from https://picsum.photos/id/237/200/300''\\
<img src="https://picsum.photos/id/237/200/300" style="width: 200px; object-fit:
contain;"></img>

\item ``<div style="width: 1024px; height: 200px; background-color: darkslategray;" />''\\
<div style="width: 1024px; height: 200px; background-color: darkslategray;" />

\item ``<input type="submit" value="Submit!"></input>'' \\
<input type="submit" value="Submit!"></input>

\item ``table that says Fruits, Apple, Lime, Blueberries in the first column and Color, Red,
Green, Blue in the second column''\\
<table style="width: 250px; border-collapse: collapse; border: 1px solid black; text-align:
center;"><tr><th>Fruits</th><th>Color</th></tr><tr><td>Apple</td><td>Red</td></tr
><tr><td>Lime</td><td>Green</td></tr><tr><td>Blueberries</td><td>Blue</td></tr>
</table>

\end{itemize}

\subsection*{S3.3. Subjects}

\subsubsection*{S3.3.1. Recruiting} Subjects were recruited from the online labor market UpWork.com. For the posting requesting programmers (``coders''), 100 subjects were recruited, and for the one requesting non-programmers (``non-coders''), 50 were recruited. 

The following postings were used to recruit subjects:

\vspace{6pt}


\textit{Non-Coders Needed For Web Experiment (Open to all with no coding background)}

\begin {displayquote}
"We are MIT researchers doing experiments on tools that help create websites. We are looking for non-coders that can help us test our interface. If you are a coder, look for our other job here \url{https://www.upwork.com/ab/applicants/1425936000678547456/job-details}.

Your job will be to replicate a couple of website mockups.
We estimate this would take 45 mins - 1.5hrs of your time for the whole job. We might also want to discuss your experience during the task to improve it.

If you perform well on this job, we will prioritize you for similar future projects.

To be a good fit for this project you should have no experience with coding (If you code, you should look and will do better by doing the “Website Mockup Replication in HTML (For Coders)” experiment).

You will receive a base pay of \$10 dollars and will receive a bonus of \$10 dollars for each problem you get correctly amounting to up to \$30 dollars for the two problems

You will complete this job using the special interface we provide. The link to the interface will be sent to you once we have accepted you for the job.

If you are interested in this project, please submit a proposal.”

\end{displayquote}


\noindent \textit{Frontend Web Developer Needed For Experiment }

\begin{displayquote}
"We are MIT researchers doing experiments on tools that help create websites. We are looking for coders with experience in HTML and CSS to test our interface.

Your job will be to replicate a couple of website mockups.
We estimate this would take 45 mins - 1.5 hrs of your time for the whole job. We may also want to discuss your experience with the task to improve it.

If you perform well on this job, we will prioritize you for similar future projects.

To be a good fit for this project you should have experience coding in HTML and CSS.

You will receive a base pay of \$10 dollars and will receive a bonus of \$10 dollars for each problem you get correctly amounting to up to \$30 dollars

You will complete this job using the special interface we provide. The link to the interface will be sent to you once we have accepted you for the job.

If you are interested in this project, please submit a proposal.”

\end{displayquote}


\subsubsection*{S3.3.2. Screening} The status of subjects as programmers or non-programmers was confirmed by using text message conversations with the subjects and by reviewing their resumes and biographical information on UpWork. The software interface used in the experiment also asked the subjects to self-classify as a programmer (``coder'') or non-programmer (``non-coder''). Subjects who did not self-classify for the group they were recruited for were not included in the study.

\subsubsection*{S3.3.3. Instructions and incentives} All subjects were provided with the following instructions:

``In this experiment, you are asked to solve two different problems. The main goal of each problem is to try and replicate a mockup website as accurately as possible. You can receive up to \$30 for participating in this experiment.
\begin{itemize}
\item The base amount will be \$10 for completing both problems.
\item You will receive an additional bonus of up to \$10 dollars per problem,
conditional on getting it right and depending on how fast you do it.
Before starting the actual problem you will have some time with a practice
mockup to get familiar with the interface, so that you don’t lose your time
in the actual problem when it counts towards the amount of bonus you
get (the faster you do it the bigger the bonus)''.
\end{itemize}

\textit{Training Session}. Subjects were encouraged to practice using a simple example web page they could try to replicate with unlimited time. The performance on this training session was not evaluated. Subjects were additionally provided with videos of how to interact with the interface.

\textit{Payment}. Each subject received a base payment of \$10 for participating in the experiment. They received no bonus for incorrect submissions. For correct submissions, their bonus was determined according to the following rule: 
\begin{itemize}
\item \$10.00 if done more than 10 minutes faster than the average
\item \$7.50 if done between the average and 10 minutes faster than the average
\item \$5.00 if done in a time between the average and 10 minutes slower than the average.
\item \$2.50 if done more than 10 minutes slower than the average
\end{itemize}
Note that this detailed payment rule was not provided to the subjects. They received only the instructions specified above.

\subsection*{S3.4. Scoring quality of code generated}

\subsubsection*{S3.4.1. Evaluating the correctness of code generated} The correctness of the code generated for each page was scored based on the total number of components in the page.  A component was defined as one individual HTML element (e.g., a button or a link), and each component was worth a maximum of 4 points, one point each for:

\begin{itemize}
\item \textit{position} relative to the overall frame of the webpage (0.5 points if the position is slightly off).
\item \textit{content}, text or graphical (0.5 points if mostly correct, e.g., if the correct text says “Over 100 locations” and the subject said “100 locations”).
\item \textit{color}, that is, text color, background color, button color, etc.
\item \textit{functionality}, that is, what the component does (i.e., text just displays text, links should link to external pages). 
\end{itemize}

Since each component was worth a maximum of 4 points, the image in Fig. S7a was worth 27x4 = 108 points, and the one in Fig. S7b was worth 13x4 = 52 points. We consider a submission “correct” if it receives at least 90\% of the maximum number of points available for it.

All the code generated was evaluated by the same person for consistency. A second person rated 10\% of the code submissions to measure inter-rater reliability. As expected, following the rubric, both raters had an almost perfect overlap never differing by more than one point out of the 52 and 108 total points per web page.

\subsubsection*{S3.4.2. Quality threshold} For accuracy, we calculate the proportion of submissions which are correct as defined above, and we expect at least 80\% of submissions in a condition to be “correct.”

\subsubsection*{S3.4.3. Level of effort} For all analyses, we excluded any subjects who didn’t put in effort, defined as spending less than 10 minutes and using less than 7 calls to GPT-3 for the “human-computer” (“HC”) condition; and as spending less than 10 minutes and generating less than 15 lines of code for the “human-only” (“H”) condition. These are reasonable numbers below which the task cannot be performed. 

These criteria led to excluding 3 out of 100 subjects in Study 2 (2 in condition HC, 1 in condition C), and 2 out of 50 subjects in Study 3.

\pagebreak
\subsection*{S3.5. Robustness Checks}

\subsubsection*{S3.5.1. Various methods of calculating confidence intervals for ratios of means}
Bonett and Price~\cite{bonett2020confidence} give several methods for calculating confidence intervals for ratios of means. In all cases, we report in the main text the results of using the method Bonett and Price recommend most, but as Table \ref{tab:CIvariants} shows, the results of using all these different methods with our data are very similar. 

\begin{table*}[ht]
 \caption{Alternative methods of calculating 95\% confidence intervals for the ratio of means (as summarized by Bonnet and Price~\cite{bonett2020confidence}). The method recommended by Bonnet and Price (last column) is reported in the main text.}
 \label{tab:CIvariants}
 \centering
 \resizebox{\textwidth}{!}{
 \begin{tabular}{lcc|ccc}
   \toprule
	 \multirow{2}{*}{Study}      &   \multirow{2}{*}{Ratio}        &
	 \multirow{2}{*}{$\hat{\rho}$}        &    \multicolumn{3}{c|}{Method} \\
    &  &   & Fieller     & Delta   & \textbf{Recommended} \\
   \midrule
   \textit{Study 2} & & & &\\
   Speed for programmers with and without GPT-3 (Paired-Samples) &  $\frac{X_{HC}}{X_H}$  & 1.17 & [1.04, 1.32] & [1.03, 1.31] & \textbf{[1.04, 1.32]}\\
   \textit{Study 3} & & & & \\
   Speed for programmers and non-programmers (Independent-Samples) & $\frac{X_{HC}}{X_{HC'}}$  & 1.07 & [0.67, 1.50] & [0.92, 1.22] & \textbf{[0.93, 1.24]}\\
   Cost for programmers with and without GPT-3 (Paired-Samples) & $\frac{C_{HC}}{C_H}$ & 0.97 & [0.88, 1.12] & [0.88, 1.12] & \textbf{[0.87, 1.11]}\\
   Cost for programmers and non-programmers (Independent-Samples) & $\frac{C_{HC}}{C_{HC'}}$ & 1.20 & [0.52, 1.94] & [0.96, 1.44] & \textbf{[0.98, 1.46]}\\
   \bottomrule
 \end{tabular}}
\end{table*}

\subsubsection*{S3.5.2. Regression for only programmers that passed}
For robustness, we run the same regression only for the population of programmers that were successful, defined as obtaining at least 90\% of the total possible points on the task. This regression has 169 observations. Table~\ref{tab:passedregression} shows the results.
The exponential of the condition coefficient remained similar improving to 1.33 with a confidence interval of [1.14, 1.55].

\begin{table}[ht]
 \caption{Study 2 Robustness Regression. Only for observations that had a score of 90\%+. The exponential of the condition coefficient is $\hat{\rho}$, the multiplicative increase from using GPT-3. 169 observations from 96 programmers.}
 \label{tab:passedregression}
 \centering
  \begin{tabular}{ccccccc}
   \toprule
	
        \multirow{2}{*}{Effect} & \multirow{2}{*}{Estimate} & \multirow{2}{*}{SE} &  \multirow{2}{*}{p} & \multirow{2}{*}{95\% CI LL}  & \multirow{2}{*}{95\% CI UL} & \multirow{2}{*}{$e^{Estimate}$} \\
   
    \\
    \midrule
   Intercept & 3.632 & 0.068 & 0.000 & 3.498 & 3.765 &37.788\\
   Condition & 0.287 & 0.078  & 0.000 & 0.134 & 0.439 & 1.332 \\
   Task & -0.030& 0.069  & 0.667 & -0.166 & 0.106& 0.970\\
   Order & -0.347& 0.060 & 0.000 & -0.465 & -0.230 &0.707 \\
   Subject RE & 0.038 & 0.073 \\

   \bottomrule
 \end{tabular}
\end{table}

\subsubsection*{S3.5.3. Removing Subject Random Effects}

We analyze a regression without random effects for the population of programmers using a standard ordinary least squares in Table \ref{tab:olsregression}. The regression has 197 observations from 100 programmers. We see that the exponential of the condition coefficient remains almost the same at 1.26 still showing significance but with a slightly bigger variance and a confidence interval of [1.06, 1.51].

\begin{table}[ht]
 \caption{Study 2 Robustness Regression.  OLS Regression without random effects. The exponential of the condition coefficient is $\hat{\rho}$, the multiplicative increase from using GPT-3. 194 Observations from 97 programmers.
}
 \label{tab:olsregression}
 \centering
 \begin{tabular}{ccccccc}
   \toprule
	
        \multirow{2}{*}{Effect}  & \multirow{2}{*}{Estimate} & \multirow{2}{*}{SE} &  \multirow{2}{*}{p} & \multirow{2}{*}{95\% CI LL}  & \multirow{2}{*}{95\% CI UL} & \multirow{2}{*}{$e^{Estimate}$} \\
   
    \\

   \midrule
   Intercept & 3.697 & 0.074  & 0.000 & 3.551 & 3.845 & 40.326\\
   Condition & 0.238 & 0.090 &  0.009 & 0.060 & 0.416 &1.269\\
   Task & -0.100 & 0.078  & 0.203 & -0.254 & 0.054 &0.904\\
   Order & -0.324 & 0.069  & 0.000 & -0.461 & -0.188 & 0.723 \\

   \bottomrule
 \end{tabular}
\end{table}

\subsection*{S3.6. Regression Derivation}

To estimate $\rho$ using a regression, we derive a generalized mixed-effects linear regression model for the special case (which occurs in our experiments) where there are two conditions and two tasks which can occur in two orders. We begin with the following multiplicative model of performance:

\vspace{6pt}
\hspace{35pt} $s_{ij}=\beta\frac{a_i}{d_j}f_{ij}g_{ij}e_{ij}$

\vspace{6pt}
\noindent where

\hspace{35pt} $s_{ij} = $ speed of subject i on task j

\hspace{35pt} $a_{i} = $ ability of subject i

\hspace{35pt} $d_{j} = $ difficulty of task j

\hspace{35pt} $f_{ij} =$ facilitation effect on performance of the condition in which subject i does task j

\hspace{35pt} $g_{ij} = $ effect on performance of the order (first or second) in which subject i does task j 

\hspace{35pt} $e_{ij} =$ error for subject i on task j 

The model says that relative to some baseline speed $\beta$, speed is increased by greater ability \textit{a} of the subject and by greater facilitation \textit{f} of the condition, and speed is decreased by greater task difficulty \textit{d}. We also assume that there is some effect \textit{g} based on whether the task occurs first or second. Finally, we assume that error is multiplicative, not additive. That is, errors are modeled as percentage increases or decreases not absolute increases or decreases.

With these assumptions, the performance of subject i on task j can be modeled as:


$$s_{ij}=\beta\bigg(\prod^{n}_{i=1}a_i^{q_i}\bigg) \times \bigg(\prod^{2}_{j=1}d_j^{-T_j}\bigg) \times \bigg(\prod^{n}_{i=1}\prod^{2}_{j=1} f_{ij}^{C_{ij} } g_{ij}^{O_{ij}}\bigg) \times e_{ij} $$

\noindent where

\hspace{35pt} $n$ is the number of subjects,

\hspace{35pt} $q_i =
\begin{cases}
      1 & \text{for subject } i\\
      0 & \text{otherwise}\\
\end{cases}$

\hspace{35pt} $T_j=
\begin{cases}
      1 & \text{for task } j\\
      0 & \text{otherwise}\\
\end{cases}$

\hspace{35pt} $C_{ij}=
\begin{cases}
      1 & \text{for the condition in which subject } i \text{ does task } j \\
      0 & \text{otherwise}\\
\end{cases}$

\hspace{35pt} $O_{ij}=
\begin{cases}
      1 & \text{for the order in which subject } i \text{ does task } j \\
      0 & \text{otherwise}\\
\end{cases}$

\vspace{6pt}
This equation can be transformed into a linear regression by treating one option for each indicator variable as part of $\beta$ and then taking natural logarithms on both sides:
$$ln \thinspace s_{ij}=ln \thinspace \beta + \sum^{n-1}_{i=1}q_i ln \thinspace a_i - T_1 \thinspace ln \thinspace d_1 + C_{11} \thinspace ln \thinspace f_{11} + 
O_{11} \thinspace ln \thinspace g_{11} + ln \thinspace e_{ij}$$

Using the observed values for $s_{ij}$ and the indicator variables for $T_j$, $C_{ij}$, and $O_{ij}$ for each observed value, we can use linear regression to estimate the values of $ln \thinspace \beta$, $ln \thinspace d_1$, $ln \thinspace f_{11}$, and $ln \thinspace g_{11}$. We can then relabel these values as $\beta_0$, $\beta_1$, $\beta_2$, and $\beta_3$.
Finally, we can take the exponentials of these logs to recover the estimated values of the original variables: $\beta$, $a_i$, $d_1$, $f_{11}$, and $g_{11}$. We estimate the subject random effects by taking $v_i=ln \thinspace a_i$.

\subsection*{S3.7. Main Regressions}

Tables~\ref{tab:mainregression} and \ref{tab:HCregression} contain the Study 2 and programmers vs non-programmers regressions, respectively.

\begin{table}[ht]
 \caption{Study 2 Regression. Comparison of programmers with and without GPT-3. The exponential of the condition coefficient is $\hat{\rho}$, the multiplicative increase from using GPT-3. 194 observations from 97 programmers.}
 \label{tab:mainregression}
 \centering
 \begin{tabular}{ccccccc}
   \toprule
	
        \multirow{2}{*}{Effect}  & \multirow{2}{*}{Estimate} & \multirow{2}{*}{SE} &  \multirow{2}{*}{p} & \multirow{2}{*}{95\% CI LL}  & \multirow{2}{*}{95\% CI UL} & \multirow{2}{*}{$e^{Estimate}$} \\
   
    \\

   \midrule
   Intercept & 3.697 & 0.066 & 0.000 & 3.567 & 3.827 & 40.326\\
   Condition & 0.240 & 0.076  & 0.002 & 0.091 & 0.390 & 1.271\\
   Task & -0.098& 0.066 & 0.137 & -0.227 & 0.031 & 0.376\\
   Order & -0.329 & 0.059  & 0.000 & -0.444 & -0.214 & 0.720\\
   Subject RE & 0.046 & 0.063 \\

   \bottomrule
 \end{tabular}
\end{table}

\begin{table}[!ht]
 \caption{Study 3 Regression. Comparison of programmers and non-programmers, both using GPT-3. The exponential of the condition coefficient is ${\rho}$, the multiplicative increase from using programmers instead of non-programmers. 193 observations from 145 subjects.}
 \label{tab:HCregression}
 \centering
 \begin{tabular}{ccccccc}
   \toprule
	
        \multirow{2}{*}{Effect}  & \multirow{2}{*}{Estimate} & \multirow{2}{*}{SE} &  \multirow{2}{*}{p} & \multirow{2}{*}{95\% CI LL}  & \multirow{2}{*}{95\% CI UL} & \multirow{2}{*}{$e^{Estimate}$} \\
   
    \\

   \midrule
   Intercept & -3.792 & 0.099 & 0.000 & -3.986 & 3.597 & 0.023\\
   Condition & 0.015 & 0.118  & 0.902 & -0.217 & 0.247 & 1.015 \\
   Task & 0.311 & 0.049 & 0.000 & 0.216 & 0.406 & 1.365\\
   Order & -0.221 & 0.048  & 0.000 & 0.126 & 0.315 & 0.802 \\
   Subject RE & 0.410 & 0.435 \\

   \bottomrule
 \end{tabular}
\end{table}

\subsection*{S3.8. Estimating costs} We compute the total cost per task per subject for the different conditions by computing the human cost which depends on the time spent in a task and the hourly rate per subject; and the GPT-3 cost based on the number of calls and the average cost per call. The Davinci engine is \$0.06 per 1000 tokens, each call uses an average of 66 tokens plus an additional 578 tokens for the prompting. The average cost per GPT-3 call is \$0.039. 

\subsection*{S3.9. Cost Regressions} Cost Regressions $HC$ vs $H$ and $HC$ vs $HC'$ (Programmers vs Non-programmers).

\begin{table}[ht]
 \caption{Study 2 Cost Regression. Comparison of programmers with and without GPT-3. The exponential of the condition coefficient is $\hat{\rho}$, the multiplicative increase from using GPT-3. 194 observations from 97 programmers.}
 \label{tab:costregression}
 \centering
 \begin{tabular}{ccccccc}
   \toprule
	
        \multirow{2}{*}{Effect}  & \multirow{2}{*}{Estimate} & \multirow{2}{*}{SE} &  \multirow{2}{*}{p} & \multirow{2}{*}{95\% CI LL}  & \multirow{2}{*}{95\% CI UL} & \multirow{2}{*}{$e^{Estimate}$}  \\
    \\
   \midrule
   Intercept & 2.507 & 0.068 & 0.000 & 2.373 & 2.640 & 12.268\\
   Condition & -0.116 & 0.072  & 0.109 & -0.258 & 0.026 & 0.890\\
   Task & -0.147 & 0.063 & 0.019 & -0.269 & -0.024 & 0.863\\
   Order & -0.311 & 0.056  & 0.000 & -0.420 & -0.201 & 0.733\\
   Subject RE & 0.196 & 0.159 \\
   \bottomrule
 \end{tabular}
\end{table}

\begin{table}[ht]
 \caption{Study 3 Cost Regression. Comparison of non-programmers and programmers with GPT-3. The exponential of the coefficient is the multiplicative increase due to the different independent variables. 193 observations from 145 subjects.}
 \label{tab:costregressionHC}
 \centering
 \begin{tabular}{ccccccc}
   \toprule
	
        \multirow{2}{*}{Effect}  & \multirow{2}{*}{Estimate} & \multirow{2}{*}{SE} &  \multirow{2}{*}{p} & \multirow{2}{*}{95\% CI LL}  & \multirow{2}{*}{95\% CI UL} & \multirow{2}{*}{$e^{Estimate}$} \\
   
    \\

   \midrule
   Intercept & 2.135 & 0.108 & 0.000 & 1.922 & 2.347 & 8.457\\
   Condition & 0.337 & 0.130  & 0.010 & 0.082 & 0.591 & 1.401 \\
   Task & -0.351 & 0.046 & 0.000 & -0.441 & -0.261 & 0.704\\
   Order & -0.199 & 0.046  & 0.000 & -0.289 & -0.110 &0.819 \\
   Subject RE & 0.512 & 0.553 \\

   \bottomrule
 \end{tabular}
\end{table}

\newpage
\subsection*{S3.10. Duration and Quality Distributions}

\begin{figure}[ht]
 \centering
  \includegraphics[width=.8\textwidth]{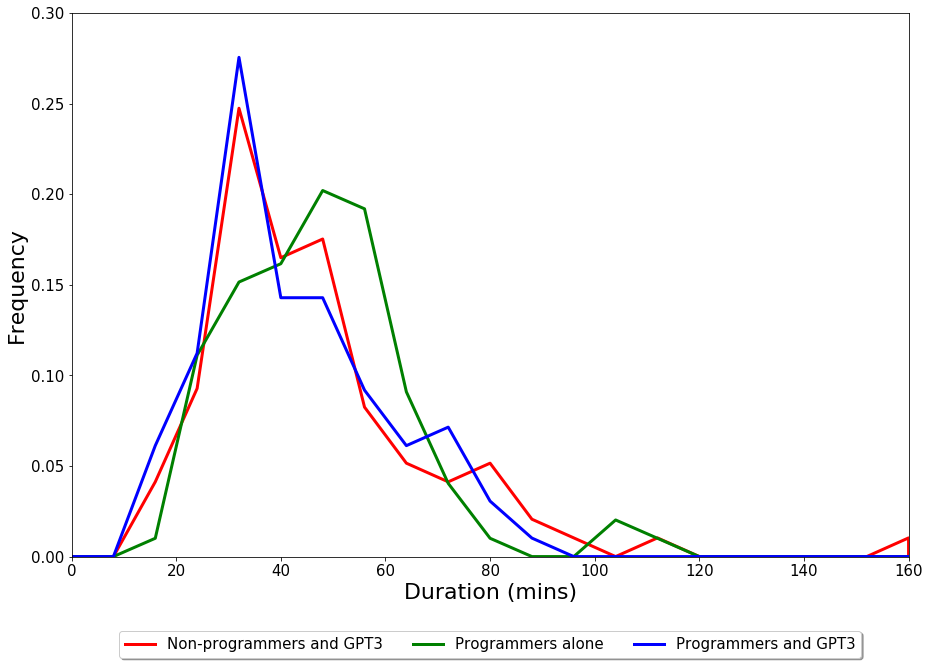}    
  \caption{Probability distribution of subjects across task duration for the three conditions.}
    \label{fig:gptPDFDuration}
\end{figure}

\begin{figure}[ht]
 \centering
  \includegraphics[width=\textwidth]{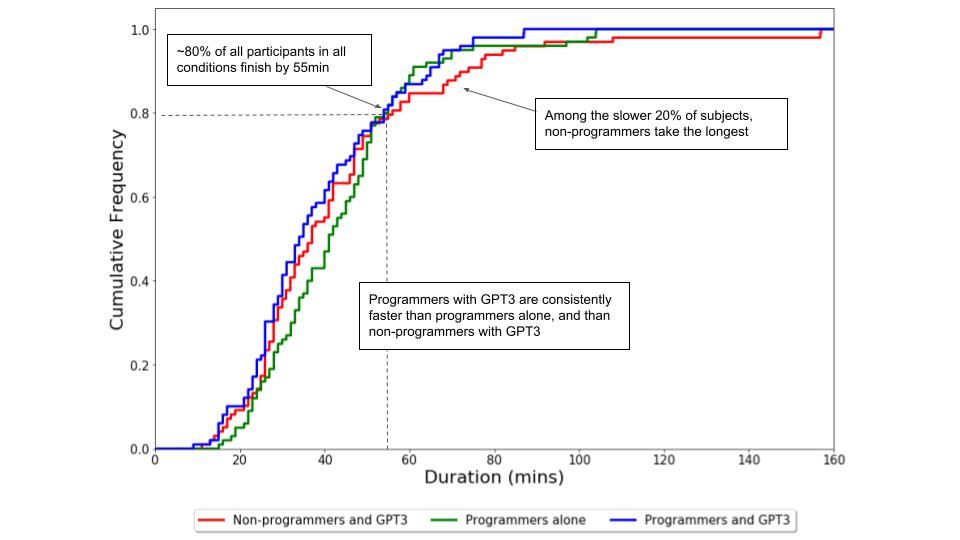}    
  \caption{Cumulative distribution of subjects across task duration for the three conditions.}
    \label{fig:gptCDFDuration}
\end{figure}

\begin{figure}[ht]
 \centering
  \includegraphics[width=.8\textwidth]{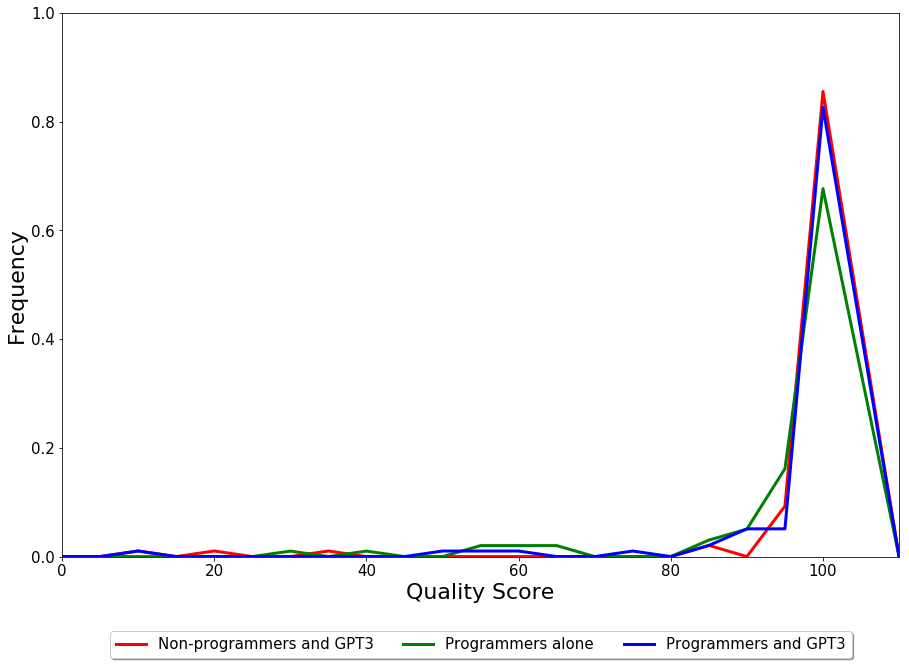}    
  \caption{Probability distribution of subjects across quality for the three conditions.  Quality is defined as percent of total points obtained on individual submissions. Almost all subjects achieved over 90\% of the possible points on their submissions.}
    \label{fig:gptPDFQuality}
\end{figure}

Figure \ref{fig:gptPDFDuration} and \ref{fig:gptCDFDuration} show, respectively, the probability distribution functions and cumulative distribution functions for task duration in the three conditions. Figure \ref{fig:gptPDFQuality} shows the probability distribution function for quality in the same three conditions.

\clearpage

\pagebreak
\bibliographystyle{ieeetr}
\bibliography{supplement}